\documentclass[a4paper, amsfonts, amssymb, amsmath, reprint, showkeys, nofootinbib, twoside,floatfix]{revtex4-2}
\usepackage[english]{babel}
\usepackage[utf8]{inputenc}
\usepackage{array}
\usepackage{float}    
\usepackage{xr-hyper} 
\usepackage[pdftex, pdftitle={Article}, pdfauthor={Author}]{hyperref}

\usepackage{epstopdf}
\usepackage{siunitx}
\usepackage{threeparttable}
\usepackage{tabularx}
\usepackage{mathtools}
\usepackage{physics}
\usepackage{xcolor}
\usepackage{graphicx}
\usepackage[left=13mm,right=13mm,top=25mm,bottom=25mm,columnsep=15pt]{geometry} 
\usepackage{adjustbox}
\usepackage{placeins}
\usepackage[T1]{fontenc}
\usepackage{lipsum}
\usepackage{csquotes}

\usepackage{fancyhdr} 
\fancyheadoffset{0cm}

\fancypagestyle{plain}{%
  \fancyhf{}%
  \fancyhead[R]{\thepage}%
}

\begin{document}
\raggedbottom

\title{Single-Shot Matrix-Matrix Multiplication Optical Tensor \\ Processor for Deep Learning}
\author{Chao Luan$^{1*}$}
\author{Ronald Davis III$^{1}$}
\author{Zaijun Chen$^{2}$}
\author{Dirk Englund$^{1}$}
\author{Ryan Hamerly$^{1,3}$}
\affiliation{$^{1}$Research Laboratory of Electronics, MIT, Cambridge, MA, 02139, USA}
\affiliation{$^{2}$Department of Electrical Engineering and Computer Science, University of California, Berkeley, CA, USA}
\affiliation{$^{3}$NTT Research Inc., PHI Laboratories, 940 Stewart Drive, Sunnyvale, CA 94085, USA}

\begin{abstract}
The ever-increasing data demand craves advancements in high-speed and energy-efficient computing hardware. Analog optical neural network (ONN) processors have emerged as a promising solution, offering benefits in bandwidth and energy consumption. However, existing ONN processors exhibit limited computational parallelism, and while certain architectures achieve high parallelism, they encounter serious scaling roadblocks for large-scale implementation. This restricts the throughput, latency, and energy efficiency advantages of ONN processors. Here, we introduce a spatial-wavelength-temporal hyper-multiplexed ONN processor that supports high data dimensionality, high computing parallelism and is feasible for large-scale implementation, and in a single time step, a three-dimensional matrix-matrix multiplication (MMM) optical tensor processor is demonstrated. Our hardware accelerates convolutional neural networks (CNNs) and deep neural networks (DNNs) through parallel matrix multiplication. We demonstrate benchmark image recognition using a CNN and a subsequently fully connected DNN in the optical domain. The network works with 292,616 weight parameters under ultra-low optical energy of $\approx$ 20 attojoules (\si{aJ}) per multiply and accumulate (MAC) at 96.4$\%$ classification accuracy. The system supports broad spectral and spatial bandwidths and is capable for large-scale demonstration, paving the way for highly efficient large-scale optical computing for next-generation deep learning.  
\end{abstract}

\maketitle
\section*{Introduction}

Advances in DNNs have revolutionized science and technology, driving innovations in computing and information processing, including nature language processing \cite{LeCun2015, vaswani2017attention}, computer vision \cite{NIPS2012_c399862d, He_2016_CVPR, Sarker2021}, multi-physics simulations \cite{He2024, RevModPhys.91.045002}, and biomedical sciences \cite{Ching2018, GOECKS202092}. To date, the exponential growth in DNN model parameters and data quantities \cite{integrated_silicon_photonics} stretches the capacity limits of conventional digital computing architectures, primarily due to the “Von Neumann” bottleneck in data movement \cite{Heuser_2020}. Though specialized computing paradigms like the Google tensor processing unit (TPU) \cite{googleedgeTPU, kumar2019scale}, field-programmable gate arrays (FPGAs) \cite{8594633,7505926}, graphics processing units (GPUs) \cite{GPU6045685, Pandey2022}, and memristor architectures \cite{10.1145/3297858.3304049, Yao2020} that merge memory and matrix multiplication computations together have been developed to improve computational throughput, reduce latency, and lower energy consumption, they are still fundamentally limited in both bandwidth and efficiency because of electronic Joule heating, capacitance, and electromagnetic crosstalk \cite{miller2017attojoule}.

Optical neural networks---which perform matrix multiplication, nonlinear activation using light---are emerging as a next-generation alternative and have the potential to address the Von Neumann bottleneck, owing to high data encoding/decoding speeds, nearly energy-free data fanout and movement, wide (100 THz) working waveband, and multiple modulation degrees (wavelength-, space-, polarization- and temporal-dimensions) of freedom \cite{shen2017deep,hamerly2019large, xu202111, wang2022optical,hamerly2021edge, zhu2022space, feldmann2021parallel, feldmann2019all, farhat1985optical, Ashtiani2022, zuo2019all, tait2017neuromorphic, shi2019deep, pappas2025reaching}. Notable advances include the demonstration of programmable nanophotonic MVM processor (PNP) using Mach–Zehnder interferometer (MZI) mesh arrays and thermo-optic phase shifters \cite{shen2017deep}, weight-bank optical MVM processors using multi-wavelength light sources and configurable ring resonators for data and weight information encoding \cite{tait2019silicon, huang2020chip}, and phase-change memory optical tensor core (PCM-OTC) for one-step MMM operations \cite{feldmann2021parallel, dong2023higher}. Recent advancements in delocalized photonic deep learning (NetCast) use optical fanout and analog time integrator for low energy MVM edge computing \cite{sludds_netcast, hamerly2019large, 10539183}.

\begin{table*}[!t]
    \centering
    \vspace{0.75em} 
    \textbf{Comparison of ONN architectures.}
    
    \vspace{0.5em} 
    \renewcommand{\arraystretch}{1.3} 
    \setlength{\tabcolsep}{3pt} 
     \begin{tabular}{@{}>{\centering\arraybackslash}p{2.5cm}|>{\centering\arraybackslash}p{4.2cm}|>{\centering\arraybackslash}p{3cm}|>{\centering\arraybackslash}p{3.1cm}|>{\centering\arraybackslash}p{4.3cm}@{}}
        \hline\hline
        \textbf{Concept*} & 
        \textbf{Hardware} & 
        \textbf{MACs/step} & 
        \textbf{MACs/step/HW} & 
        \textbf{Caveats} \\ 
        \hline
        PNP~\cite{shen2017deep} & $N^2$ MZI & $N^2$ [1-step MV] & $1/$MZI & MZI errors, BW limits. \\  
        WB~\cite{tait2017neuromorphic} & $N^2$ ring & $N^2$ [1-step MV] & $1/$ring & Ring stabilization. \\ 
        HD-ONN~\cite{hamerly2019large,chen2023deep} & $N$ ring, $N^2$ det & $N^2$ [N-step MM] & $N/$ring, $1/$det & Coherence, cyl. optics. \\ 
        NetCast~\cite{sludds_netcast} & $1$ mod, $1$ WDM & $N$ [N-step MV] & $N/$mod, $N/$WDM & Weight server. \\ 
        PCM-OTC~\cite{feldmann2021parallel} & $N^2$ PCM, $N^2+2N$ WDM$^{\dagger}$& $N^3$ [1-step MM] & $N/$PCM, $N/\text{WDM}^{\S}$ & Combiner loss, WDMs. \\ 
        \hline
        PNP+WDM & $N^2$ MZI, $N$ WDM & $N^3$ [1-step MM] & $N/$MZI, $N^2/$WDM & MZI BW, WDMs. \\ 
        WB+WDM & $N^2$ ring, $N$ WDM & $N^3$ [1-step MM] & $N/$ring, $N^2/$WDM & Ring variations, WDMs. \\ 
        HD+WDM & $N^2$ ring, $N^2$ det & $N^3$ [1-step MM] & $N/$ring, $N/$det & Coherence nightmare. \\ 
        \hline
        \textbf{This work} & $2N^2$ MZI, 1 grating & $N^3$ [1-step MM] & $N/$MZI, $N^3/$grating & Dispersive grating parallelism, crosstalk.\\
        \hline\hline
    \end{tabular}
    \vspace{0.5em} 
     \caption{Comparison of existing ONN architectures, variants of these architectures, and this work.
        The PNP~\cite{shen2017deep}, WB~\cite{tait2017neuromorphic}, 
        HD-ONN~\cite{hamerly2019large,chen2023deep}, NetCast~\cite{sludds_netcast} architectures have been experimentally demonstrated but exhibit low parallelism. Proposed enhanced variants, including PNP+WDM, WB+WDM, and HD+WDM, theoretically support high-parallelism operations but face practical limitations in large-scale feasibility and have not been experimentally demonstrated. The PCM-OTC architecture has experimentally demonstrated with high parallelism; however, it encounters serious scaling challenges (wavelength count, WDM count, and crossbar fan-in) for large-scale implementation. In contrast, our proposed architecture achieves high parallelism without being restricted by such scaling limitations.
         *PNP: Programmable nanophotonic processor. WB: Weight bank. HD-ONN: Homodyne-ONN. PCM-OTC: Phase-change memory optical tensor core. $^{\dagger}$The system needs two types of WDM, the first type WDM has $N^2$ channels, PCM-OTC architecture needs $N$ of them, the second type WDM has $N$ channels, and PCM-OTC architecture needs $N^2+N$ of them. \textsuperscript{\S}Only consider the $N$ channel WDMs.}
         \label{tab:onn_comparison}
\end{table*}

The inherent high modulation degree of freedom of light makes ONN processors specifically suitable for parallel matrix multiplications, that is, tensor operations in deep learning. The tensor operations support parallel matrix multiplications on large batches of data across multiple physical dimensions, allowing for better throughput, latency, and energy efficiency than ordinary MVMs and VVMs \cite{jouppi2017datacenter, googleedgeTPU,xu2022high}. The tensor processors are used to accelerate advanced network architectures like CNNs (multi-channel data for 3D convolutions with multiple kernels) \cite{xu2022high, feldmann2021parallel}, recurrent neural networks (RNNs) (time-series data with batch and sequence dimensions) \cite{novikov2015tensorizing}, and transformers (multi-head attention mechanisms requiring high-dimensional tensor manipulation) \cite{googleedgeTPU}. In addition, the tensor operations reduce the software implementation complexity and allow for frameworks (like TensorFlow, PyTorch) to natively support higher-dimensional data and simplify the model deployment \cite{paszke2019pytorch}. Most importantly, tensor processors support efficient memory usage and minimize the data-movement, system latency, and energy consumption \cite{jouppi2017datacenter}. These architectural benefits have motivated using optical tensor processors for high-throughput and energy-efficient optical computing, however, the realization of large-scale high-parallelism optical tensor processors remains challenging.

Table~\ref{tab:onn_comparison} summarizes the existing ONN architectures in terms of hardware (number of components), gross throughput [MAC/step], and normalized throughput (optical parallelism) [MAC/step/HW component]. The latter figure is most important since it is directly proportional to the throughput density (the scaling factor being the component size). As this comparison illustrates, current leading ONN architectures do not harness the full advantages of optics. The PNP, WB are natively single-wavelength and exploring WDM for extra parallelism introduces fundamental challenges. The PNP, which is based on MZIs, is highly sensitive to wavelength if directional couplers are used, leading in practice to a sub-nanometer window of operation so using a wide range of wavelengths can compromise accuracy; likewise, the WB only operates at designated wavelengths; On the other extreme, PCM-OTC utilizes WDM and has $O{(N)}$ parallelism per channel, but its scalability is rather limited, due to the power consuming with beam combination, the $O{(N^2+2N)}$ WDM components (which increases the system complexity), and the $O{(N^2)}$ comb lines needed to avoid crosstalk between frequency modes, so the matrix size is limited by the bandwidth of on-chip sources. In addition, the PCM has limited states, so the bit precision is low \cite{dong2023higher}. The only design that can achieve high parallelism without fundamental scaling roadblocks is the HD-ONN \cite{hamerly2019large}; however, practical issues of coherence and cylindrical optic aberrations ultimately limit its performance.

Here, we propose and demonstrate a dispersion-based, three-dimensional, high parallelism, highly scalable and low loss ONN tensor processor utilizing time-wavelength-spatial hyper-multiplexed data encoding and a parallel dispersive free space grating beam routing to improve the throughput, energy efficiency, optical sensitivity and reduce the latency. It achieves high component parallelism like the PCM-OTC and HD-WDM-ONN, while not relying on optical interference or cylindrical imaging like the HD-WDM-ONN or suffering from the fan-in losses, low bit precision, and large system complexity due to a large amount of WDMs of the PCM-OTC crossbar, so it's not limited by fundamental scaling roadblocks. The parallel dispersive ONN architecture uses spatial-division multiplexing (SDM) and WDM for “energy-free” data fanout and movement, a dispersive blazed grating for large scale, parallel, low crosstalk and low-loss beam routing, spatial-wavelength integrating and analog time integrating for high parallelism $O{(N^3)}$, high energy efficiency, high optical sensitivity and high scalability MACs. As proof of concept, the system is used to perform accurate single-shot and time-integrated parallel batch image convolutions by encoding one \(2\times 2\) kernel and one \(3\times 3\) kernel into the weight modulators and two images into the activation modulators simultaneously. We also compile a benchmark one-layer CNN using four \(2\times 2\) kernels and a subsequently fully connected DNN containing one hidden layer with 100 neurons using the optical hardware. The inference network operates with 292,616 pre-trained weight parameters and shows high throughput density, high energy efficiency, and high optical sensitivity.

\begin{figure*}[t!]
\centering
\includegraphics[width=0.7\textwidth]{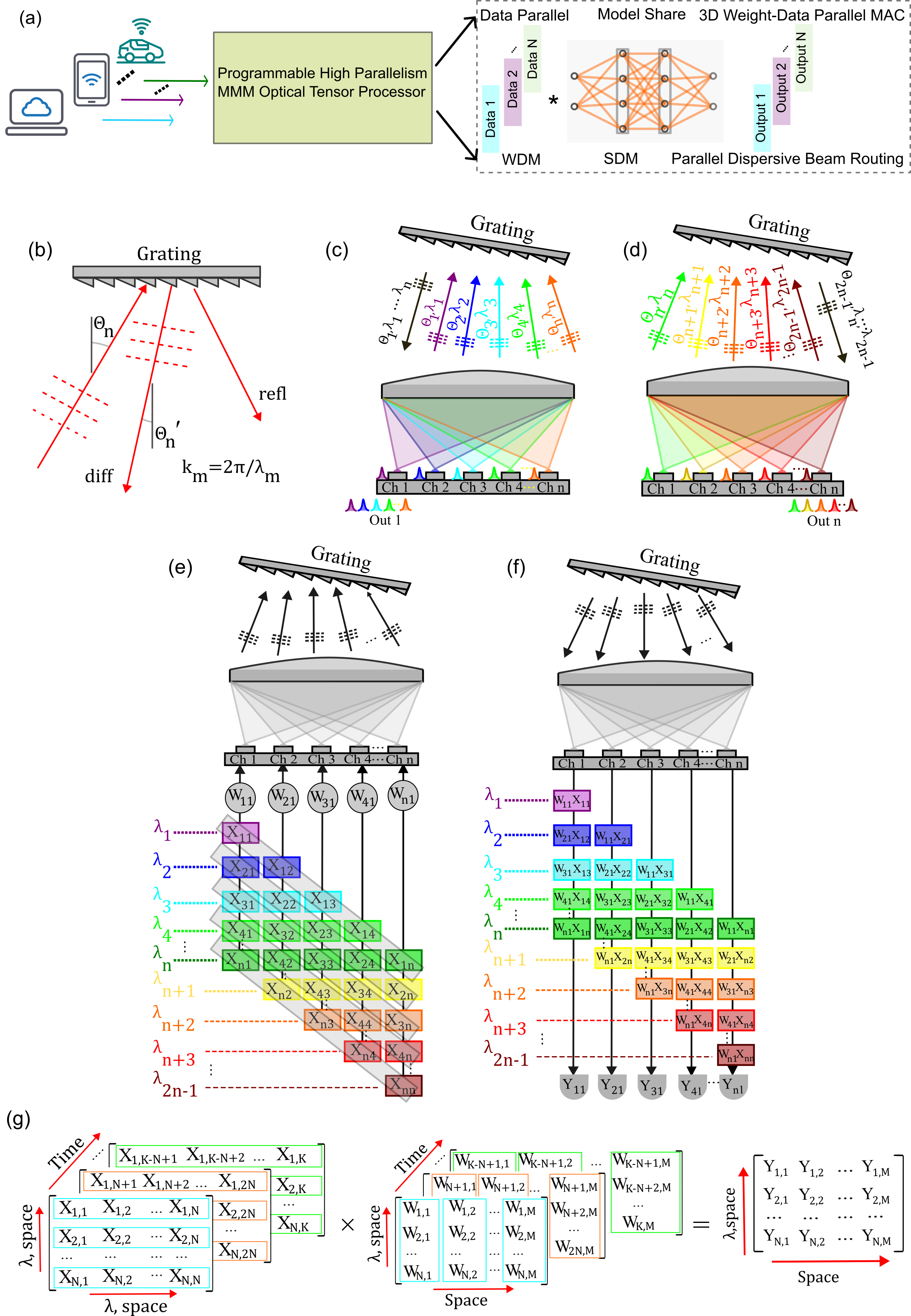}
\caption{Overview and working principle of the single-shot MMM optical tensor processor. (a) Overview of the proposed high parallelism single-shot MMM optical tensor processor, the processor supports parallel batch data input and allows for parallel weight model sharing among all the input data, performing parallel MMM. (b) Diffracted and reflected beams resulting from light incident on a dispersive grating. (c) and (d) A two-dimensional cross-section of the three-dimensional WDM combiner realized by grating diffraction off a broadband wavelength light source injected from different spatial channels. (e) and (f) A two-dimensional view of the parallel dispersive ONN architecture that supports single-shot MVM. In this scheme, the dispersive grating works as three-dimensional parallel multiplexer and demultiplexer arrays. (g) Concept layout of the space-wavelength-time multiplicated parallel MMM.}
\label{fig:1}
\end{figure*}

\section*{Working Principle and Data Architecture}
Figure~\ref{fig:1} illustrates the overview and principle of the time-wavelength-spatial hyper-multiplexed parallel optical tensor processor. The ONN processor is based on the parallel physical light dispersion law of a grating beam routing system. The optical processor utilizes WDM, SDM, parallel dispersive beam routing, and analog time integration, it supports parallel batch data inputs and allows for parallel cost-free weight model sharing among all these input data, demonstrating parallel 3D MMM with just one dispersive grating. The key effect that underlies this ONN tensor processor is the programmable chromatic dispersion from diffraction off a grating (Fig.~\ref{fig:1}(a)). A broadband plane wave incident on a grating is split into a reflected component and several diffracted orders (the grating period here is fine enough that only the first diffraction order is allowed). For input angle \(\theta\), light is diffracted at angle \(\theta'\) given by the equation: 
\begin{equation}
\label{eq1}
k\sin(\theta') = k_g - k\sin(\theta)
\end{equation}
where \(k=\frac{2\pi}{\lambda}\) and \(k_g =\frac{2\pi}{L}\) are the optical and grating wavenumbers. The effect of grating dispersion is to route the light between inputs and outputs through:
\begin{equation}
\label{eq2}
\sin(\theta_n')=\frac{\lambda_m}{L}-\sin(\theta_n)
\end{equation}
Define the following grid of wavelengths $\lambda_m$ and angles $\theta_n$:

\begin{equation}
\lambda_m = \lambda_0 + m L \Delta\theta \cos(\theta_0),
\label{eq3}
\end{equation}
\begin{equation}
\sin(\theta_n) = \sin(\theta_0) + n \Delta\theta \cos(\theta_0),
\label{eq4}
\end{equation}

where \((\theta_0,\lambda_0)\) is directly back-scattered by the grating. Eq.~(\ref{eq2}) is satisfied when:
\begin{equation}\label{eq5}
n'=m-n
\end{equation}
Eq.~(\ref{eq5}) indicates that the spatial channels are reordered in a wavelength-dependent manner. In such a system, the light emits from an array of emitters, each has different wavelengths, the emitters’ wavelength spacing and spatial pitch follow Eqs.~(\ref{eq3}-\ref{eq5}) so all the grating scattered lights are multiplexed into the same output channel. This effect is used to create a passive WDM (Fig.~\ref{fig:1}(b-c)). Fig.~\ref{fig:1}(b-c) only shows a 2D cross-section of the grating here; in practice, a vertically-coupled free-space grating is 3D and provides an array of WDMs.

Let each input channel consist of \(N\) wavelengths rather than just one (Fig.~\ref{fig:1}(d-e)). The optical intensities in these channels encode the activation matrix \(X_{i,j}\) in a position-frequency basis through programming the grating dispersion, another broadband intensity modulator is employed to encode the weight vector elements \(W_{j}\). The cascaded modulation leads to multiplication in the laser intensity. After the diffraction off the free-space grating, the intensity in each output channel is reordered. In practice, the free-space grating is 3D; it works as parallel 3D arrays of demultiplexers and multiplexers, so the weight matrix \(W_{j,q}\) could be defined in a 3D structure in a single time step, as indicated in Fig.~\ref{fig:1}(g). Amplified photodetectors are used to read out the computing results, and in a single time step, the system yields the 3D matrix-matrix product of:
\begin{equation}
Y_{i,q}=\sum\nolimits_{j} X_{i,j} * W_{j,q}\label{eq6}
\end{equation}

To efficiently use the spatial-wavelength interleaving nature of the grating, the wavelength number in each spatial channel should equal the total spatial channel numbers of the activation modulators. The system throughput is \(N^2M\) MACs in a single time step through wavelength and spatial integration over \(N\times M\) output channels while only needing \(2N-1\) different wavelengths. Time multiplexing is incorporated to extend the inner dimension through accumulating the photovoltages over \(T\) time steps using the analog time-integrating receivers \cite{sludds_netcast, hamerly2019large}. Through the time integration, the output of the optical tensor processor is:


{\small
\begin{equation*}
Y_{i,q} 
= \sum\Bigl(\sum\nolimits_{j=1}^N X_{i,j} * W_{j,q} \;\cdots\; 
             \sum\nolimits_{j=K-N+1}^{K} X_{i,j} * W_{j,q}\Bigr)
\end{equation*}
\vspace{-0.6em}
\begin{equation}
= \sum\nolimits_{j=1}^{K} X_{i,j} * W_{j,q}
\label{eq:7}
\end{equation}
}

\section*{Single-Shot MMM Optical Tensor Processor Setup }
\begin{figure*}[t!]
\centering
\includegraphics[width=0.7\textwidth]{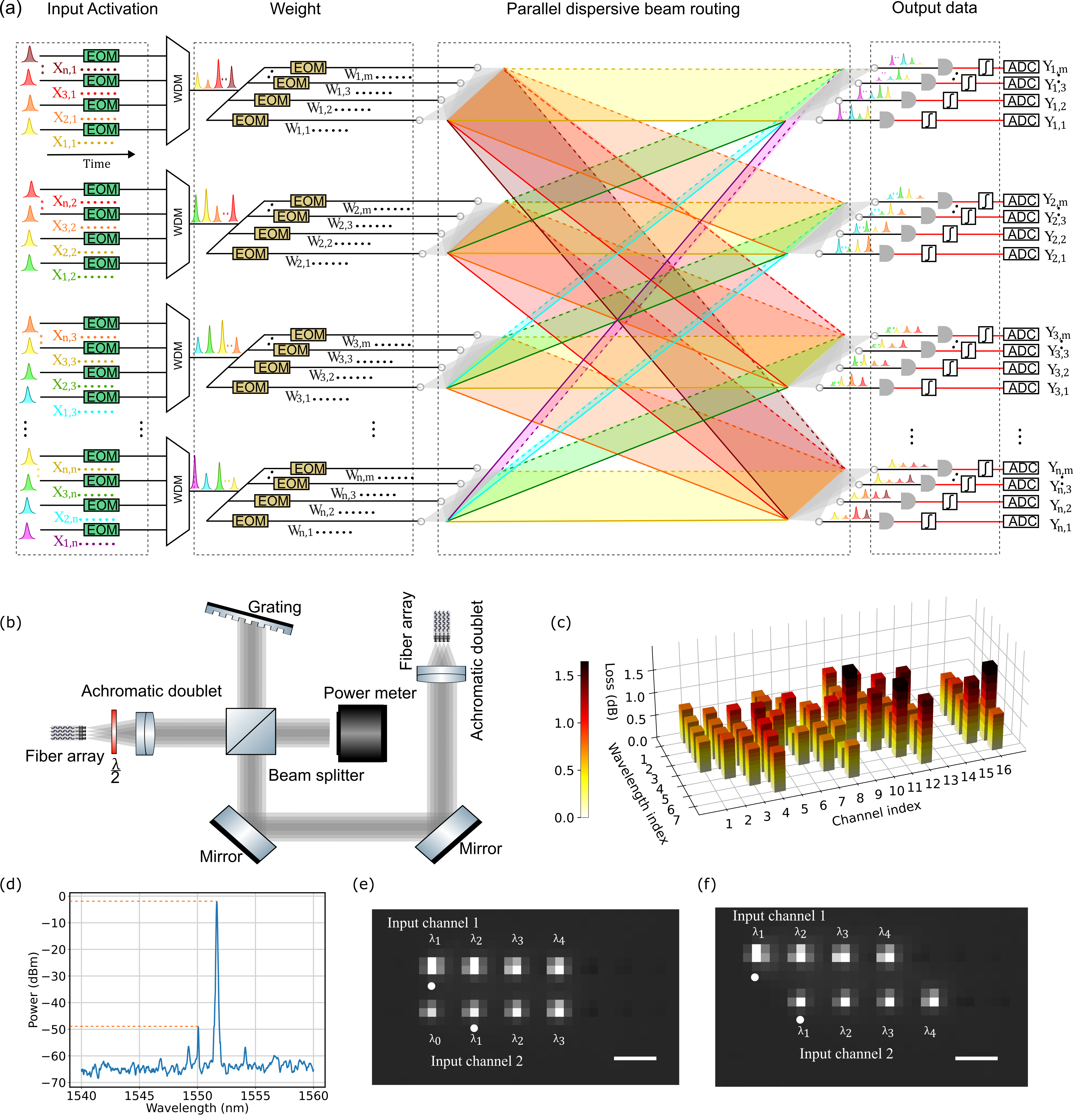}
\caption{Architecture of the ONN processor and characterization results of the parallel dispersive grating beam routing. (a)  Schematic of the parallel MMM optical tensor processor, \(N^2\) intensity modulators from \(N\) groups with \(2N-1\) different wavelengths work as the signal modulator and \( NM\) intensity modulators work as the weight modulators, the parallel dispersive grating routing system reorders the MAC operation according to the input wavelength and spatial channel and in a single time step the MMM is demonstrated with \(N\) MACs per channel, analog time integrating detector extended the system to parallel MMM. (b) Schematic of the parallel dispersive grating beam routing system. (c) Loss characterization of the parallel dispersive grating beam routing system. (d) Crosstalk of the parallel dispersive grating beam routing system. (e) and (f) Parallelism characterization of the dispersive grating beam routing system, scale bar is 150\,\textmu m }
\label{fig:2}
\end{figure*}

Fig.~\ref{fig:2} shows the sketch of the single-shot MMM optical tensor processor. The processor uses a ``weight-stationary'' frame \cite{8114708, 7993626}. The data matrix \(X_{N\times K}\) is mapped in \(T = K/N\) time steps to a bank of \(N^2\) individually addressable LiNbO\textsubscript{3} modulators. These \(N^2\) data modulators are divided into \(N\) groups (Fig.~\ref{fig:2}(a)), each group contains \(N\) modulators, and each modulator encodes data onto a separate wavelength, with equal wavelength spacing. Each data modulator group defines one column of $X$, so the \(N\times N\) data sub-matrix is defined in a single time step. To align with the spatial pitch of the fiber array feeding into and out from the parallel dispersive grating, the wavelengths for adjacent modulator groups are shifted with one wavelength spacing (see Fig.~\ref{fig:1}(b-d)).

The optical outputs of each data-modulator group are wavelength-multiplexed and then fanned out to \(M\) different spatial channels for free data movement. Each fanout channel light now contains all the \(N\) different wavelengths and being modulated by only one broadband intensity weight modulator. The weight matrix \(W_{K\times M}\) is again encoded in \(T = K/N\) time steps by \(N\times M\) LiNbO\textsubscript{3} intensity modulators, so that in a single time step, the \(N\times M\) weight submatrix is defined. 
The modulated \(N\times M\) channel outputs, each containing \(N\) different wavelengths, are then guided to the parallel dispersive grating beam routing system through an \(N\times M\) channel single mode fiber array. The 3D parallel dispersive grating beam routing system works as a parallel \( N^2\times M\)-channel input, \(N^2\times M\)-channel output multiplexers and demultiplexers, reordering the weight matrix and data matrix according to Eqs.~(\ref{eq3}-\ref{eq5}). As a consequence, in a single clock cycle, the data matrix with the same row and the weight matrix with the same column are multiplied and accumulated through programmable parallel WDM and SDM, achieving a throughput of \(N^2 \times M\) MACs/cycle. The time-integrating receivers extend the inner product dimension through accumulating these \(N\) wavelength and spatial dimension matrix multiplication results over \(T = K/N\) time steps, generating the tensor product $Y_{i,q}$ (Eq.~(\ref{eq:7})).
\begin{figure*}[t!]
\centering
\includegraphics[width=0.7\textwidth]{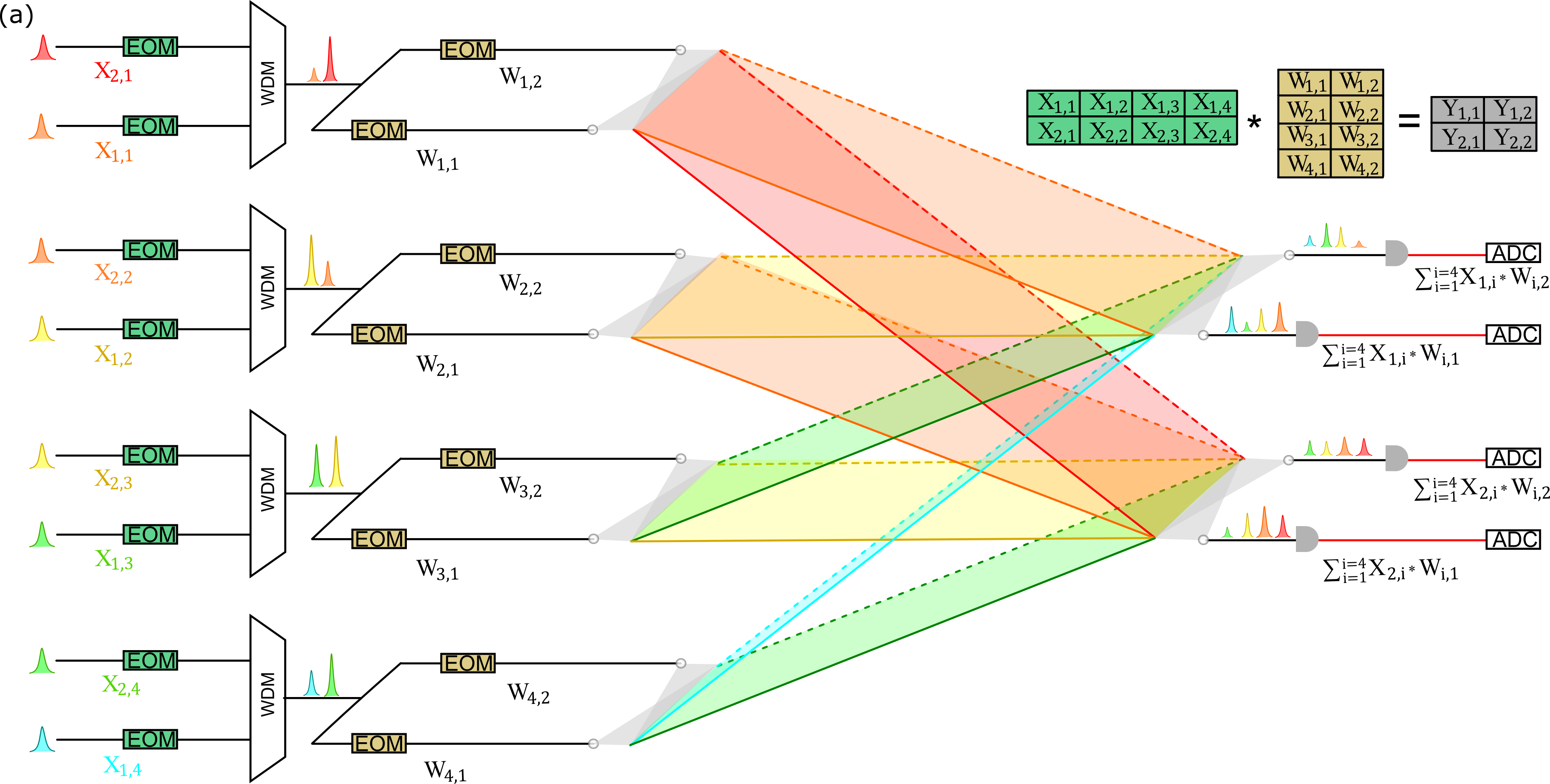}
\vspace{20mm}
\includegraphics[width=0.7\textwidth]{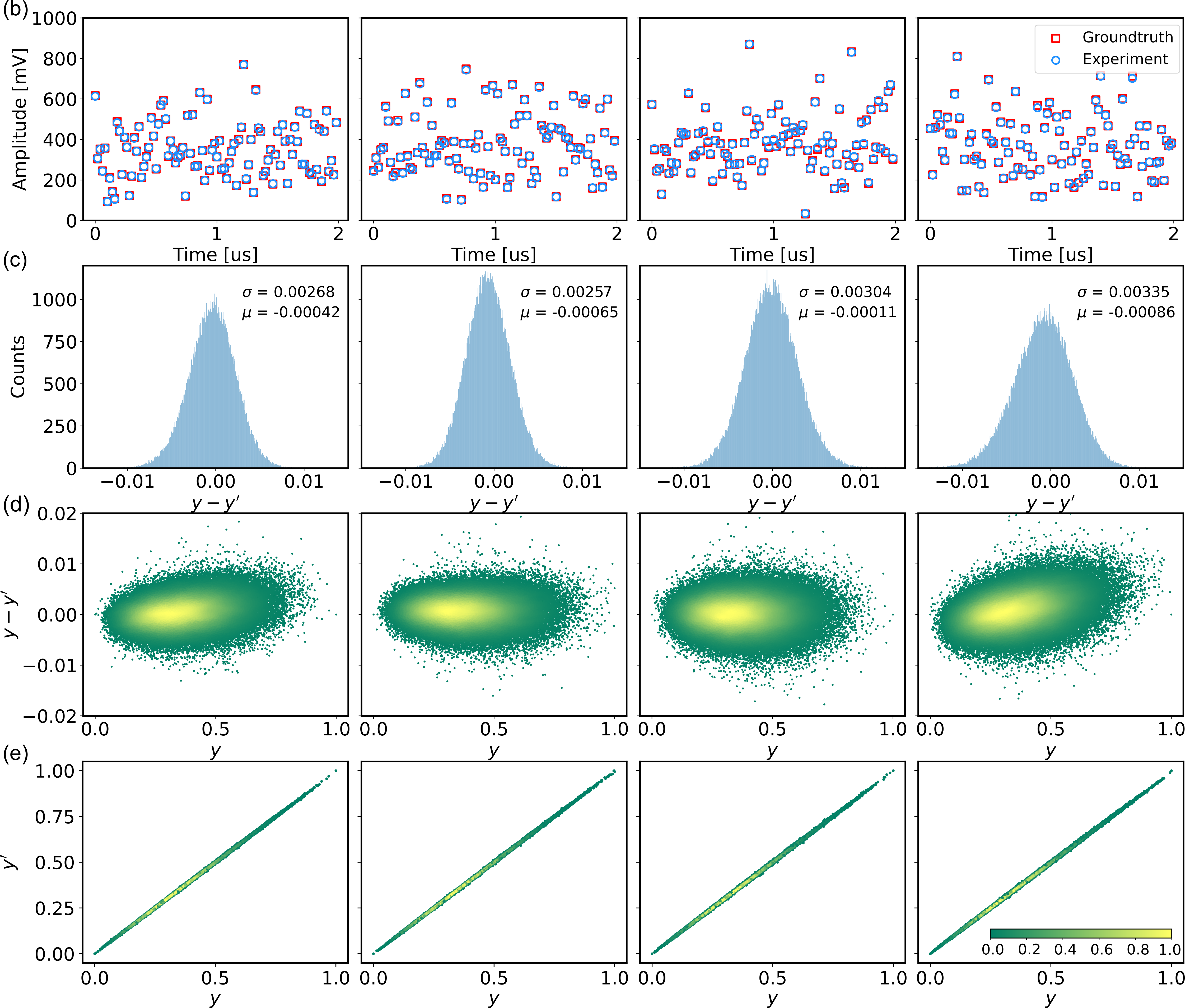}

\caption{Experimental setup and verification of the signal-shot \(2\times 4\times 2\) MMM processor computing accuracy across multiple wavelength and spatial channels. (a) \(2\times 4\times 2\) MMM optical tensor processor setup, four data modulator groups, each with two intensity modulators encodes the \(2\times 4\) data matrix and four weight modulator groups encode the \(4\times 2\) weight matrix in a single time step, through the grating beam routing system, a single time step \(2\times 4\times 2\) MMM is demonstrated among four different channels. (b)-(e) Time trace of the measured and expected MMM processor among these four different channels. (f)-(i) Experiment-theory difference standard deviation distribution of these four channels, theory analog results are designated as \(y\) and optical computing results are designated as \(y'\). All the measured standard deviation is around 0.003, indicating over 8 bits of calculation accuracy. (j)-(m) Distribution of the theory and experimental results among four different channels.}
\label{fig:3}
\end{figure*}

\section*{Parallel dispersive grating beam routing system characterization}

\begin{figure*}[t!]
\centering
\includegraphics[width=0.9\textwidth]{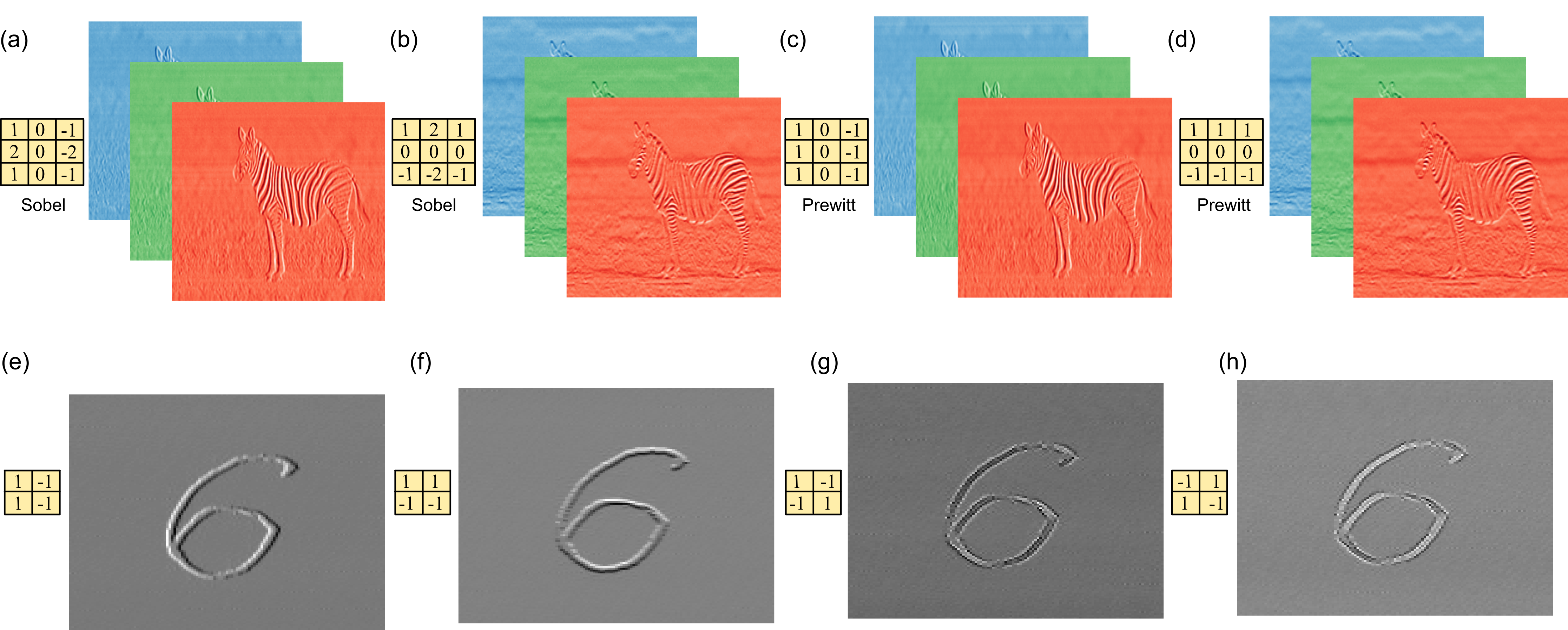}
\caption{Parallel convolution operation using the architecture. (a)-(d) A 3D convolution of a colored image with four different \(3\times 3\) kernels are demonstrated by mapping the 9 kernel elements of each kernel into 4 different spatial channels among 3 different time steps and recorded by analog time integrators, through the convolution, horizontal and vertical features are successfully extracted. (e)-(h) A 2D convolution of a gray image with four \(2\times 2\) kernels, for each kernel, the four elements are mapped to four different spatial channels, and the convolutional results are recorded by amplified photodetectors.}
\label{fig:4}
\end{figure*}

Fig.~\ref{fig:2}(b) shows a 2D cross-section of the 3D grating beam routing system. We measure the grating beam routing system loss by comparing the receiver fiber array output power with the corresponding input fiber array input power, under both single wavelength and multiple wavelength input conditions. For both cases, the grating beam routing system loss comes from three parts: the 3-dB beam splitter loss, the grating diffraction loss, and the free space to the receiver fiber array coupling loss. In the quantization, we omit the 3-dB beam splitter loss because a polarization-sensitive beam splitter and a Faraday rotator can circumvent it. To reduce the coupling loss, the \(4\times 4\) input fiber array is single mode (core diameter of 10\,\textmu m and pitch of 150\,\textmu m), and the \(4\times 4\) receiver fiber array is multimode (core diameter of 50\,\textmu m and pitch of 150\,\textmu m). Based on these conditions, the parallel dispersive grating beam routing system loss is summarized in Fig.~\ref{fig:2}(c). The system loss is comparable with the commercial single channel WDM loss but provides \(O(N^2M)\) scale parallelism advantage, only one parallel dispersive grating is needed, which otherwise needs 32 commercial WDMs to achieve the same throughput (supplementary material section III.F), in addition, unlike commercial WDM systems that have fixed channel wavelengths, our beam-routing system provides programmable channel wavelengths by adjusting the incident angle on the dispersive grating.

Crosstalk of the grating beam routing system is low.  Crosstalk is a critical parameter as it will influence the matrix multiplication computing signal-to-noise ratio (SNR) and bit precision and determines the minimal wavelength spacing of the system. We measure the system crosstalk by sending four different wavelengths (equally spaced, with 200~GHz spacing) into one input channel and measure the output spectrum power at different receiver channels (150\,\textmu m spatial spacing). Through the parallel dispersive grating routing, the input wavelengths are distributed among four different receiver channels, with negligible channel (-46 dB) crosstalk, which likely originates from undesired scattering from the fiber array front surfaces. We perform the crosstalk measurement for all the fiber array input channels, and the measured values are all around -50 dB, as shown in Fig.~\ref{fig:2}(d). The 200~GHz wavelength spacing is determined by the bandwidth of the available WDM in our lab, which is used to multiplex different data modulator wavelengths. The wavelength spacing could be further reduced, meantime keeping a low grating crosstalk and high SNR (supplementary material section III.D).

The grating output beams are focused after the receiver achromatic doublets and recorded by using an infrared camera (Goldeye G-008, 30\,\textmu m pixel size). The focused output beams are programmably distributed according to Eqs.~(\ref{eq3}-\ref{eq5}). For example, when we inject two broadband sources with identical wavelengths into adjacent input channels, they emerge at adjacent positions (one fiber pitch apart) in the output plane (Fig.~\ref{fig:2}(f)). By contrast, if the two sources differ by one wavelength channel spacing, they emerge from the same output channel (Fig.~\ref{fig:2}(e)). The optical parallelism of the current grating routing system is 64 (\(N^2 \times M\)) MACs/cycle.

\section*{Demonstration of the Single-shot \(2\times 4\times 2\) MMM}

 The grating beam routing system supports single-shot MMM. Fig.~\ref{fig:3}(a) shows the experimental setup for the single-shot \(2\times 4\times 2\) optical MMM processor based on the parallel dispersive grating beam routing scheme. Limited by the available arbitrary waveform generator (AWG, Keysight M3202A, 12 bits, 400 MHz, 16 channels), the maximum E/O modulator numbers we can drive simultaneously is 16. We therefore constructed a single-shot \( 2\times 4\times 2\) MMM optical tensor processor with 16 E/O modulators. A $2\times 4$ data matrix and a $4\times 2$ weight matrix, each containing 30,000 randomly generated positive analog entries, are encoded onto the carrier light intensity through LiNbO\textsubscript{3} modulators. These modulators exhibit high extinction ratios ($>30$\,dB), wide 3\,dB bandwidths (20\,GHz), 4.5\,dB insertion loss, and a drive voltage $V_\pi$ of 4.5\,V. 
 
 We perform analog values MAC operation directly in the optical domain. To do this, the analog values are linearly mapped to the photodetector output voltages and the corresponding modulator input drive voltages are inversely extracted by solving modulators’ voltage transfer functions (supplementary material section II.C). To keep the modulator in the quasi-linear region, the D.C.\ bias voltages are set at the modulator output power transfer function quadrature point. A multi-channel DC source (Qontrol Q8iv) together with a homemade in-time automatic bias controller algorithm compensates for drift in these 16 modulators (Supplementary Section~II.E), ensuring stable transfer functions and obviating repeated calibrations (Supplementary Section~II.C). The 16-channel AWG with a peak-peak voltage of 3 V drives these modulators, the AWG output RF waveform is pulse amplitude modulated and the clock rate \( C\) is downsampled from \(1\,\mathrm{GSas^{-1}}\) to \(50\,\mathrm{MSas^{-1}}\). The AWG output RF signals are timed to compensate for the system optical path difference, so all the detected optical signals are synchronized. We use eight fiber polarization controllers to unify the polarization across modulators, and a free-space half-wave plate to optimize the grating diffraction efficiency for all the eight input channels. 
 The \(2\times 4\times 2\) optical MMM processor output results are measured by four amplified photodetectors and read out by a 16-channel high precision waveform digitizer (ATS9416, 14 bits, 65 MHz), these four output channels are characterized in parallel. Thanks to the real-time bias controller, the modulator intensity transfer function remains stable, and the experimental time traces agree closely with the theoretical predictions (Fig.~\ref{fig:3}(b-e)). Comparing experiment to groundtruth, we calculate the residuals of these four output channels, and the measured standard deviation distribution is 0.003. This corresponds to a bit precision beyond 8\, bits (ENOB=6.6 bits), confirming both low crosstalk and high parallelism in the parallel dispersive grating beam routing system. On the other hand, because of the wavelength integration nature of the grating, the processor throughput is 4 MACs per channel per clock cycle, this increases the SNR of the output signal through accumulating over four channel signals but with only one \(O/E\) conversion noise and one ADC read out noise. The same measurements are performed by shifting the input laser wavelengths over 2 THz and shifting the input spatial channels, the results show high reproducibility and high consistency, indicating the broad spectrum bandwidth and broad spatial bandwidth of the system.

\begin{figure*}[t!]
\centering
\includegraphics[width=1\textwidth]{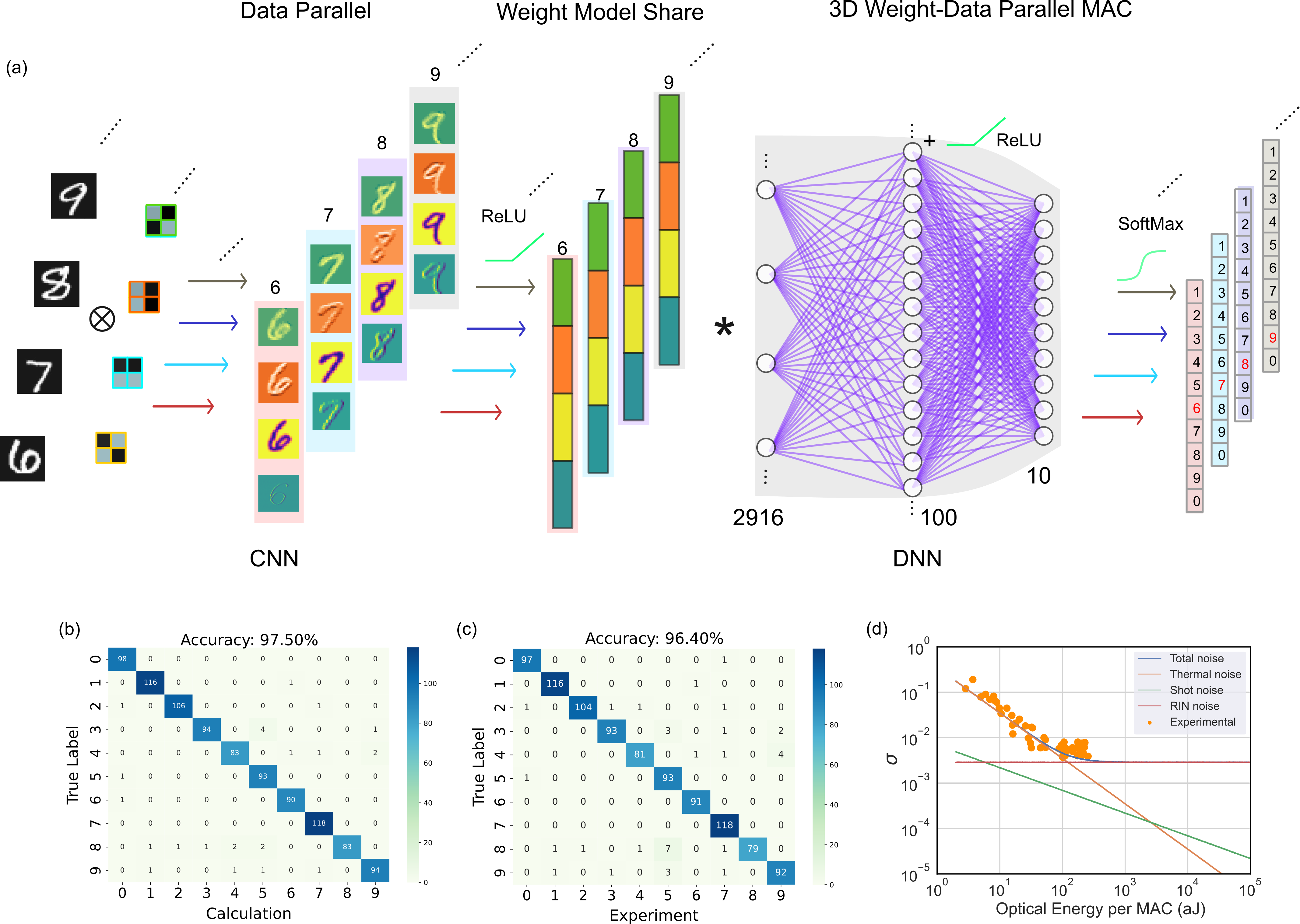}
\caption{Using parallel MMM optical tensor processor to perform benchmark MNIST image classification. (a) The architecture of the network, the kernel and weight matrix are mapped to the weight modulators and the inputs are maped to the data modulators. The ONN processor supports parallel batch data input and allows for parallel weight model sharing for all the input data. We perform parallel CNNs and following fully connected DNNs. (b) and (c) Calculated and measured confusion matrix of the system and (d) Optical sensitivity of the architecture to perform the image classification, the system has a high sensitivity even at (\(\approx 20\,\mathrm{aJ}\)). As indicated, the thermal noise and relative intensity noise (RIN) dominate over shot noise for the parallel dispersive grating beam routing ONN system.}
\label{fig:5}
\end{figure*}

\section*{Demonstration of the Parallel image Convolutional, benchmark CNNs and DNNs for image classification}
Having demonstrated the basic capabilities to perform single-shot MMM optical computing by using our parallel dispersive grating beam routing system, we now show, in Fig.~\ref{fig:4}, experimental examples of optical tensor operation of processing parallel convolutions by using batch data images and multiple kernel matrixes at the same time. Two kernel matrices, one \(2\times 2\) and the other \(3\times 3\), are mapped into the weight modulators, and two data image matrices, with sizes matching the convolution kernels, are mapped into the data modulators simultaneously. The convolutional outputs for the \(2\times 2\) kernel are recorded by two amplified photodetectors while the convolutional outputs for the \(3\times 3\) kernel are recorded by two analog time integrating receivers. 

For the \(2\times 2\) kernel, the four elements of each kernel are mapped to four weight modulators in four different wavelengths, so each modulator has a constant drive voltage, while for the \(3\times 3\) kernel, the nine elements of each kernel are mapped to four modulators in three time steps. By using the time integrating detectors, we can achieve even larger kernels (supplementary material section IV.B). The kernel matrices contain both positive values and negative values, they are mapped into the modulators in different time rounds (supplementary material section IV.A). The convolutions highlight different edges that can clearly be seen, for example, Fig.~\ref{fig:4}(a) and (c) emphasize horizontal edges, whereas Fig.~\ref{fig:4}(b) and (d) highlight vertical edges, confirming the high parallelism and low crosstalk of the grating routing system.

Now we use this parallel dispersive grating beam routing optical processor for Modified National Institute of Standards and Technology (MNIST) image recognition. We perform a benchmark one-layer CNN and a following fully connected DNN inferences using the optical processor. Fig.~\ref{fig:5}(a) shows the architecture of the network. We apply four \(2\times 2\) kernels to extract the features of these images, each input image, with \(28\times 28\) pixels, is encoded in 729 time-steps to four data modulators (supplementary material section IV.B).
The grating beam routing optical processor performs parallel batch operations, two \(2\times 2\) kernels, and two MNIST images are convolved by the grating beam routing processor simultaneously. The outputs from the convolutional layer, which contain 2,916 pixels of 1,000 images, are detected by amplified photodetectors. A series of operations like decoding the detector data from voltages to analog values, ReLU nonlinear activations, and data normalizations are performed digitally. The CNN outputs are fed into a fully connected DNNs, implemented by the same optical system, which consists of two layers ($2916 \times 100$ and $100 \times 10$), a ReLU activation, and a SoftMax operation on the output.

Fig.~\ref{fig:5}(b-c) shows the confusion matrix of the network output, where we demonstrate 96.4$\%$ computation accuracy of our optical processor for 1,000 MNIST images, which is close to the pre-trained (fully digital) accuracy of the network (97.5$\%$). As performed before, we verify the system's spatial and wavelength bandwidth and scalability by shifting the input laser wavelength and the grating input channel and achieve the same classification accuracy. This result shows the potential for this architecture to support high-wavelength bandwidth and high-spatial parallelism in real-world deployed systems using conventional components.

\section*{System performance}
The spatial-wavelength-temporal hyper-multiplexing and parallel dispersive grating beam routing enable optical tensor processors with high parallelism, low energy consumption, high optical sensitivity, and large network scalability.   

For a \(N\times M\) channel fiber array, the system parallelism is \( N^2M\). For the current scheme, \(N=M=4\), and \(C=50\,\mathrm{MSas^{-1}}\), the system throughput is \( 2N^2MC=0.0064\)~TOPS. The throughput is currently limited by the available AWG clock rate \(C\) and the fiber array channel numbers \(N\times M\). This limitation can be addressed in the near-term with a high-speed AWG and large channel fiber arrays. 

Especially, the throughput scales cubically with the fiber array channels, and the \(N\times M\) fiber array channels could be further expanded. The increase of \(N\) is losses as it is based on WDM of the input activation modulator wavelengths. The channel number \(N\) is limited by the chromatic aberration of the lens under broadband wavelengths. We use achromatic doublet lenses to reduce the aberration as it provides better broadband and off-axis performance than aspheric lenses. On the other hand, the multimode receiver fiber array has a much larger Rayleigh length and a much larger tolerance for the aberration and divergence of the input beam than normal single-mode fiber. The high-resolution grating supports input light wavelength spacing of 50~GHz (supplementary material section III.D) meanwhile keeping low crosstalk between adjacent fiber array channels. The maximum allowed working wavelength range is over 50 $\mathrm{nm}$ (supplementary material section III.E), this corresponds to a wavelength channel number \(2N-1\) of 125 and fiber array channel number \(N\) over 60. The expansion of \(M\) is based on the optical spatial fanout and the loss is proportional to the fanout channel numbers. The use of wavelength integration and time integrating receiver lowers the required optical power per readout as the ADC read-out signal SNR is determined by the read-out power instead of the power per MAC, and as a result, the grating beam routing system has a much larger tolerance for the fanout loss of the 1 over \(M\) channel splitter, in addition, the channel number \(M\) fanout is not restricted by the lens chromatic aberration as the grating WDM only happens in the horizontal (N) direction. All these make it possible for the implementation of large-scale fiber arrays to increase the system parallelism and system throughput. For example, a \(30\times 30\) channel fiber array---which needs 900 data modulators, 900 weight modulators, and 59 different wavelengths---at optical clock rates of \( C=10\,\mathrm{GSas^{-1}} \) would allow a system parallelism of 27,000 and a throughput of 540 TOPS (supplementary material section III.E).

The parallel dispersive optical tensor processor is designed to minimize power usage using optical fanout, parallel beam routing, and analog time integrating receiver \cite{hamerly2019large}. One weight modulator is used to encode parameters across \(N\) data modulators. In addition, these \(N\) data modulators are fanout to \(M\) weight modulators by SDM, which otherwise need \(NM\) data modulators. The system needs \(N^2+NM\) modulators, though the energy costs of these individual components are high, the system has high parallelism, performing \(N^2M\) MACs of work per time step, allowing for \(P_{E/O}\left(\frac{1}{N} + \frac{1}{M} \right)\,\mathrm{pJ}\) per MAC where \(P_{E/O} \) is the single shot energy consumption of one modulator. If not using the fanout, the system would need \(2N^2M\) modulators and 2~pJ per MAC. The system also needs \(N\times M\) photodetectors, performing \(N\) MACs of work per time step per channel, allowing for \(P_{O/E}\frac{1}{N} \,\mathrm{pJ}\) per MAC where \(P_{O/E}\) is the single shot energy consumption of one photodetector. This also applies for the \(N^2+NM\) channel DAC, allowing for \(P_{DAC}\left(\frac{1}{N}+\frac{1}{M}\right)\,\mathrm{pJ}\) per MAC where \(P_{DAC}\) is the energy consumption of one AWG channel to perform one-time digital to analog conversion.

On the other hand, the analog time integration accumulates the wavelength multiplexed MAC information over \(T\) time steps among \(NM\) channels. The system MACs is \(TN^2M\) and only needs one-time ADC readout, so the readout energy is only \(P_{ADC}\frac{1}{TN}\,\mathrm{pJ}\) per MAC per channel. Assuming a \(30\times 30\) channel fiber array and a time integrating detector accumulation period of 100, the total system energy consumption is 130 \(\mathrm{fJ}\) per MAC. Simple changes to the modulators, such as making use of the chip scale thin film LiNbO\textsubscript{3} modulator (\(\approx 100\,\mathrm{fJ}\) per MAC), would enable \(\approx 75\,\mathrm{fJ}\) per MAC system energy efficiency. Segment optical DAC, which uses binary drive voltage and has a low energy consumption of \(P_{ODAC}\)\(\approx 40\,\mathrm{fJ}\) per operation at 8 bits output precision \cite{moazeni201740}, could avoid the use of high energy consumption and high bit precision electro DAC and decrease the system energy consumption to 3 \(\mathrm{fJ}\) per MAC.

We show the energy sensitivity of the system (Fig.~\ref{fig:5}(d)) by measuring the MAC accuracy at different optical energies. The system shows high precision at a low optical energy of 20~aJ per MAC. We achieve this through accumulation over \(N\) wavelengths in a large \(T\) time step and readout only once. The time-integrating receiver accumulates \(T\) time steps before the readout, allowing only one ADC readout noise and \(T\) times O/E conversion noise per measurement among \(NT\) MACs. Compared with normal amplified photodetector without time integration and wavelength integration, the readout noise is \(\sqrt{NT}\) times smaller and the O/E conversion noise is \(\sqrt{N}\) times smaller. This reduced noise level of our system also gives benefits of reduced needed photodetector optical power as less \(\sqrt{NT}\) signal electron is required. 

The optical sensitivity could be further increased by using a large channel fiber array through multiplexing more wavelengths or using a time integrator with a smaller capacitor \cite{sludds_netcast}. Notably, the wavelength multiplexing and time integration will both accumulate among multiple multiplications and then accumulate only once, however, the wavelength multiplication accumulates in the optical domain and only one O/E conversion noise is added among \(N\) MACs, while the time integration accumulates the multiplication over \(T\) time steps and in each step an O/E conversion noise is added, in addition, the analog time integrator capacitor leakage will induce noise to the data.

The latency of the current architecture is mainly limited by the analog time integrator, which accumulates \(T\) time steps over \(N\) different wavelengths before the readout. Currently, analog time integration is the only method to perform accurate fully connected DNN with large weight matrix size in the optical domain. Our parallel dispersive grating beam routing ONN processor uses multi-dimensional hyper-multiplexing parallel data encoding, model sharing, and single-step 3D activation-weight connection to reduce the latency. We consider the MNIST images (each with \(28\times 28\) pixels) classifications network used in this paper. For the current \(4\times 4\) fiber array, the time to process 1,000 MNIST images is \SI{96}{\milli\second}. The latency could be reduced by multiplexing more wavelength and spatial channels through a large channel number fiber array, a \(30\times 30\) channel fiber array with \(10\,\mathrm{GSas^{-1}}\) would reduce the processing time of 1,000 images to \SI{1140}{\nano\second}.

\begin{figure}[t!]
\centering
\includegraphics[width=0.9\columnwidth]{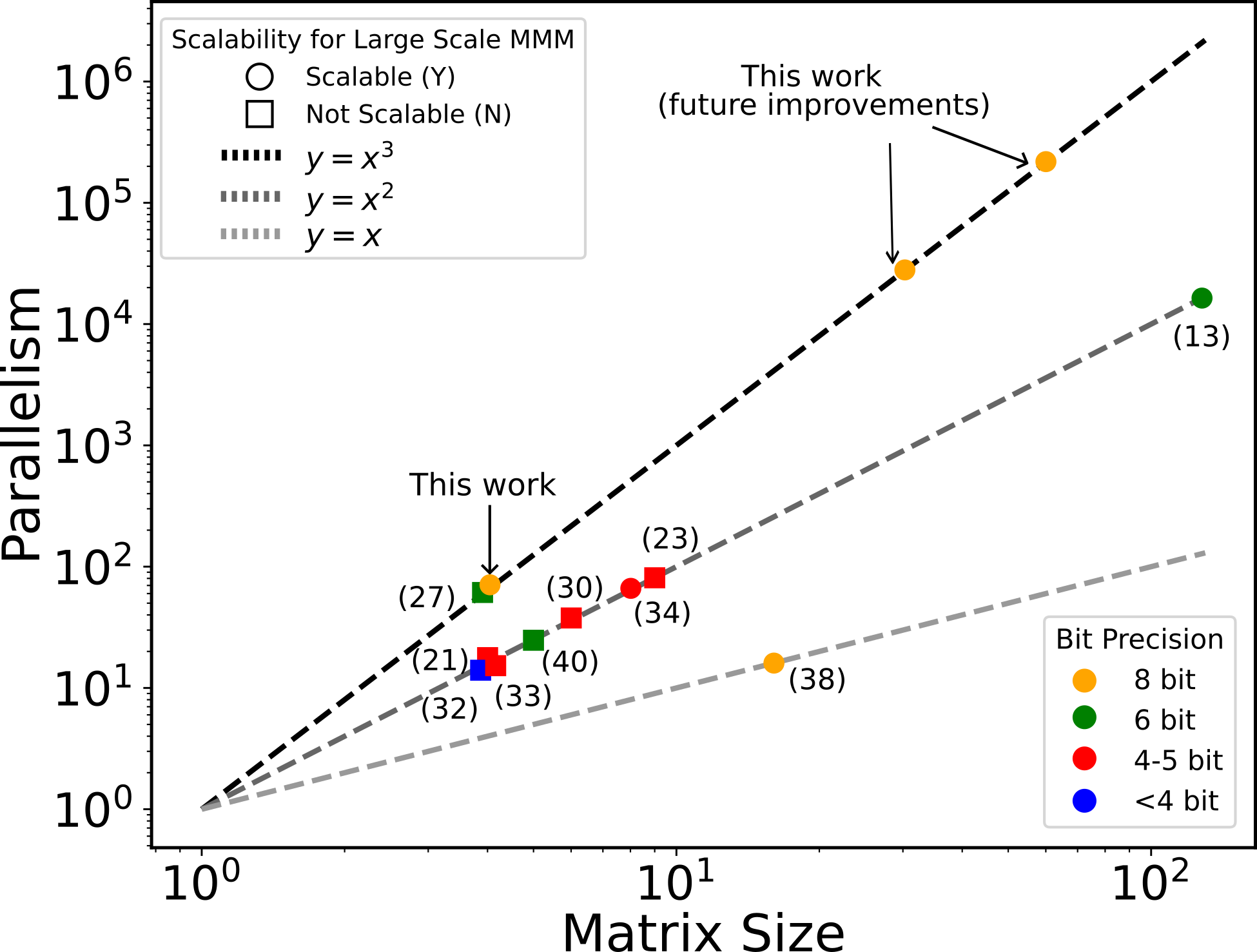}
\caption{Comparison of SoA computing systems \cite{shen2017deep, tait2017neuromorphic,sludds_netcast, feldmann2021parallel,chen2023deep, Ashtiani2022,xu202111,shi2019deep, pappas2025reaching,kumar2019scale} with matrix size and output parallelism. Our work demonstrated a wavelength-spatial-temporal hyper-multiplexing single-shot MMM ONN system by investigating the high parallelism dispersive beam routing. It follows the best parallelism tendency with high bit precision  while not limited by fundamental roadblocks when scaling to large-scale MMM operation.}
\label{fig:6}
\end{figure}

\section*{Discussion and outlook }

We have proposed and experimentally demonstrated an ONN tensor processor capable of implementing single-shot MMM using a high parallelism dispersive beam routing with the high-dimensional space-wavelength-temporal hyper-multiplexed data; the system is robust and capable of large scalability implementation. By verifying the feasibility of using grating as an array of parallel multi-channel input, multi-channel output multiplexers and demultiplexers, we provide an ONN processor with high parallelism, high throughput, high optical sensitivity, and large network scalability. We configured our system to perform single-shot \(2\times 4\times2\) MMM and a synchronous convolution of two images with two kernels. The results show high reproducibility and high consistency among different wavelength and spatial channels, indicating the broad spectrum bandwidth and spatial bandwidth of the architecture. Although these are achieved using a small-size \(4\times4\) fiber array, larger-size fiber array is envisioned for better computing density, efficiency, latency, optical sensitivity, and more general applications. 

Crucially, the parallelism of $N^2 \times M = 64$ is not an upper limit, a \(30\times 30\) or \(60\times 60\) channel fiber array is possible, with this, an overall parallelism of 27,000 or 216,000. As indicated in Fig.~\ref{fig:6}, where our architecture follows the best parallelism tendency and doesn't get limited by fundamental scaling roadblocks to large-scale parallel MMM operation. In our proof-of-concept neural network, we demonstrated high classification accuracy on the MNIST datasets with 292,616 weight parameters under ultra-low optical energy of about \(20\,\mathrm{aJ}\) per MAC at 96.4$\%$ classification accuracy. 

In summary, we have presented a robust single-shot hardware-efficient MMM optical tensor processor and demonstrated high parallelism, high throughput, low energy consumption, high optical sensitivity, and high accuracy optical inference network. The architecture has broad wavelength bandwidth and space bandwidth, opening new possibilities for larger AI models in language processing and machine vision.

\section*{Acknowledgements}
The author thanks Dr. Alex Sludds from Lightmatter for the useful discussion on the analog time integrator, Dr. Franco Wong from MIT RLE for the kind offer of the EO-space modulator and the waveshapers, and Dr. Liane Berstein from QuEra for the useful discussion on the training algorithm. 

\subsection*{Funding}
This work was funded by the research collaboration agreements with Nippon Telegraph and Telephone (NTT) research. 

\subsection*{Contributions}
Chao Luan perfected the idea, created the experiment setup, wrote the code and algorithm, conducted the experiment, performed the data analysis, and wrote the manuscript. Ronald Davis assisted with AWG programming code writing, modulator transfer function parameters extraction code writing, and preliminary digital training code writing. Zaijun Chen assisted with the free space grating setup, including initial equipment purchases, and discussed the setup. Ryan Hamerly came up with the original idea, helped debug the experiment setup, the results, and the data analysis, coordinated the equipment, and supervised the project. Dirk Englund supervised the project. Ryan Hamerly and Dirk Englund revised the manuscript. 

\section*{Competing Interests}
The authors Chao Luan, Dirk Englund, and Ryan Hamerly disclose that they are inventors on pending patent US Application No. 63/768502 where MIT is the patent applicant, which covers the Single-Shot Matrix-Matrix Multiplication Optical Tensor Processor architectures described in this work.

\section*{Data and Materials Availability}

The data from this work is stored in the server computer (QPG210212) at MIT RLE and will be made available upon reasonable request.

\section*{Code Availability}

The code used to curve fit the hardware and train the offline DNN from this work will be made available upon reasonable request.

\bibliographystyle{apsrev4-2}

\clearpage
\onecolumngrid 

\begin{center}
    {\large \bfseries Supplementary Material: Single-Shot Matrix-Matrix Multiplication Optical Tensor Processor for Deep Learning}
    
    \vspace{0.5em}
    Chao Luan$^{1*}$, Ronald Davis III$^{1}$, Zaijun Chen$^{2}$, Dirk Englund$^{1}$, Ryan Hamerly$^{1,3}$
    
    \vspace{0.5em}
    \textit{$^{1}$Research Laboratory of Electronics, MIT, Cambridge, MA, 02139, USA\\
    $^{2}$Department of Electrical Engineering and Computer Science, University of California, Berkeley, CA, USA\\
    $^{3}$NTT Research Inc., PHI Laboratories, 940 Stewart Drive, Sunnyvale, CA 94085, USA}
\end{center}


\section{Detailed comparison of existing optical neural network architectures}
\label{sec:s11}
Table~\ref{tab:onn_comparison} compares six existing ONN schemes in terms of their hardware (number of components), gross throughput [MAC/step], and normalized throughput (i.e. optical parallelism) [MAC/step/HW component]. The latter figure is most important since it is directly proportional to throughput density (the scaling factor being the component size).
The PNP \cite{shen2017deep} and weight-bank \cite{tait2017neuromorphic} schemes both involve relatively small components (MZIs and rings), but require ${N^2}$/ them to do a single-step matrix-vector product. Therefore, the normalized parallelism is 1/MZI for the PNP and 1/ring for the weight-bank scheme. WDM is problematic in both cases: the PNP, which is based on MZIs, is highly sensitive to wavelength if directional couplers are used, leading in practice to a sub-nanometer window of operation (MMIs would improve this but suffer higher losses and back scattering); likewise, the weight bank only operates at designated wavelengths. In any case, wavelength multiplexing the PNP or weight-bank system would add ${N}$ WDMs with a parallelism factor of ${N}$/WDM; this eats up a lot of chip area for passives alone.
The Homodyne-ONN \cite{hamerly2019large} was specifically optimized to maximize performance per unit area. In matrix-matrix form, it consists of $O{(N)}$ modulators, free-space fan-out optics, and $O{(N^2)}$ detectors and performs a matrix-matrix product in ${N}$ steps. Only a single wavelength is used, so the modulators can be compact rings; even so, the parallelism factor is ${N}$/ring, a factor of ${N}$ larger than the weight bank. The detectors only have a parallelism factor of unity; however, detectors can be made very small and tightly packed, i.e. a camera screen. The biggest challenge with this scheme is its use of coherence and cylindrical free-space optics, which make it uniquely sensitive to vibrations and aberrations. (Note that we could achieve ${N}$/det if we really wanted to by using WDM to compute the matrix-matrix product in a single step, but this requires $O{(N^2)}$ modulators and becomes a phase stabilization nightmare). NetCast \cite{sludds_netcast} also scores reasonably well on the metric if only the client is considered, but this analysis is only valid in the edge computing case, which is its target application. Also, note that the modulator must be a broadband MZM, which takes up much more space than the HD-ONN modulators.
Finally, the so-called optical tensor core \cite{feldmann2021parallel} combines a PCM crossbar array with WDM to achieve very high compute density, at least within the crossbar. This allows it to perform a matrix-matrix product in a single time step, achieving a parallelism of ${N}$/PCM. Like the WDM variants of the other schemes, it also requires $O{(N^2+2N)}$ WDMs, which takes up a lot of area if on-chip. There is also a very fundamental scaling problem associated with the beam combiners used in the crossbar circuit. Since the combiners are not wavelength-selective, there is a net $O{(N)}$ optical power loss from the source to the detector; this waste of light will lead to unacceptably large optical powers that ruin the energy efficiency of the design. Moreover, the  $95\%$ of input power scattered as stray light may interfere with the actual signal. These problems did not arise in the \( 4*4\) demonstration in the paper, but they will inevitably make scaling this scheme difficult.
As this comparison illustrates, most leading ONN designs do not harness the full power of photonics. The PNP, WB are natively single-wavelength and adding wavelength multiplexing introduces some fundamental challenges; as a result the parallelism of the components themselves is only $O{(1)}$. On the other extreme, the PCM-OTC has $O{(N)}$ parallelism, but is hampered by $O{(N)}$ loss in the beam combiners and the need for $O{(N^2+2N)}$ on-chip WDMs. The only design that can achieve high parallelism without fundamental scaling roadblocks is the HD-ONN; however, practical issues of coherence and cylindrical optic aberrations may ultimately limit its performance.
This introduces our high parallelism dispersive beam routing optical neural network, a new ONN based on the chromatic dispersion of free-space diffraction gratings. Grating neural network could (1) achieve high component parallelism like the PCM-OTC and HD-WDM-ONN, while (2) not relying on optical interference or cylindrical imaging like the HD-WDM-ONN or suffering from the fan-in losses of the PCM-OTC crossbar. The grating neural network realizing single-shot hardware-efficient large-scale matrix-matrix multiplication in a relatively robust optical setup.

\begin{table*}[!t]
    \centering
    \vspace{0.75em} 
    \textbf{Comparison of ONN architectures.}
    
    \vspace{0.5em} 
    \renewcommand{\arraystretch}{1.3} 
    \setlength{\tabcolsep}{3pt} 
     \begin{tabular}{@{}>{\centering\arraybackslash}p{2.5cm}|>{\centering\arraybackslash}p{4.2cm}|>{\centering\arraybackslash}p{3cm}|>{\centering\arraybackslash}p{3.1cm}|>{\centering\arraybackslash}p{4.3cm}@{}}
        \hline\hline
        \textbf{Concept*} & 
        \textbf{Hardware} & 
        \textbf{MACs/step} & 
        \textbf{MACs/step/HW} & 
        \textbf{Caveats} \\ 
        \hline
        PNP~\cite{shen2017deep} & $N^2$ MZI & $N^2$ [1-step MV] & $1/$MZI & MZI errors, BW limits. \\  
        WB~\cite{tait2017neuromorphic} & $N^2$ ring & $N^2$ [1-step MV] & $1/$ring & Ring stabilization. \\ 
        HD-ONN~\cite{hamerly2019large,chen2023deep} & $N$ ring, $N^2$ det & $N^2$ [N-step MM] & $N/$ring, $1/$det & Coherence, cyl. optics. \\ 
        NetCast~\cite{sludds_netcast} & $1$ mod, $1$ WDM & $N$ [N-step MV] & $N/$mod, $N/$WDM & Weight server. \\ 
        PCM-OTC~\cite{feldmann2021parallel} & $N^2$ PCM, $N^2+2N$ WDM$^{\dagger}$& $N^3$ [1-step MM] & $N/$PCM, $N/\text{WDM}^{\S}$ & Combiner loss, WDMs. \\ 
        \hline
        PNP+WDM & $N^2$ MZI, $N$ WDM & $N^3$ [1-step MM] & $N/$MZI, $N^2/$WDM & MZI BW, WDMs. \\ 
        WB+WDM & $N^2$ ring, $N$ WDM & $N^3$ [1-step MM] & $N/$ring, $N^2/$WDM & Ring variations, WDMs. \\ 
        HD+WDM & $N^2$ ring, $N^2$ det & $N^3$ [1-step MM] & $N/$ring, $N/$det & Coherence nightmare. \\ 
        \hline
        \textbf{This work} & $2N^2$ MZI, 1 grating & $N^3$ [1-step MM] & $N/$MZI, $N^3/$grating & Dispersive grating parallelism, crosstalk.\\
        \hline\hline
    \end{tabular}
    \vspace{0.5em} 
     \caption{Comparison of existing ONN architectures, variants of these architectures, and this work.
        The PNP, WB, HD-ONN, NetCast architectures have been experimentally demonstrated but exhibit low parallelism. Proposed enhanced variants, including PNP+WDM, WB+WDM, and HD+WDM, theoretically support high-parallelism operations but face practical limitations in large-scale feasibility and have not been experimentally demonstrated. The PCM-OTC architecture has experimentally demonstrated with high parallelism; however, it encounters serious scaling challenges (wavelength count, WDM count, and crossbar fan-in) for large-scale implementation. In contrast, our proposed architecture achieves high parallelism without being restricted by such scaling limitations.
         *PNP: Programmable nanophotonic processor. WB: Weight bank. HD-ONN: Homodyne-ONN. PCM-OTC: Phase-change memory optical tensor core. $^{\dagger}$The system needs two types of WDM, the first type WDM has $N^2$ channels, PCM-OTC architecture needs $N$ of them, the second type WDM has $N$ channels, and PCM-OTC architecture needs $N^2+N$ of them. \textsuperscript{\S}Only consider the $N$ channel WDMs.}
         \label{tab:onn_comparison}
\end{table*}

\newpage
\section{Perform multiplication and accumulation using optics}
\subsection{Subsection 1: E/O modulator characterization }
\label{sec:s21}
\begin{figure}[H]
    \centering
    \includegraphics[width=0.3\textwidth]{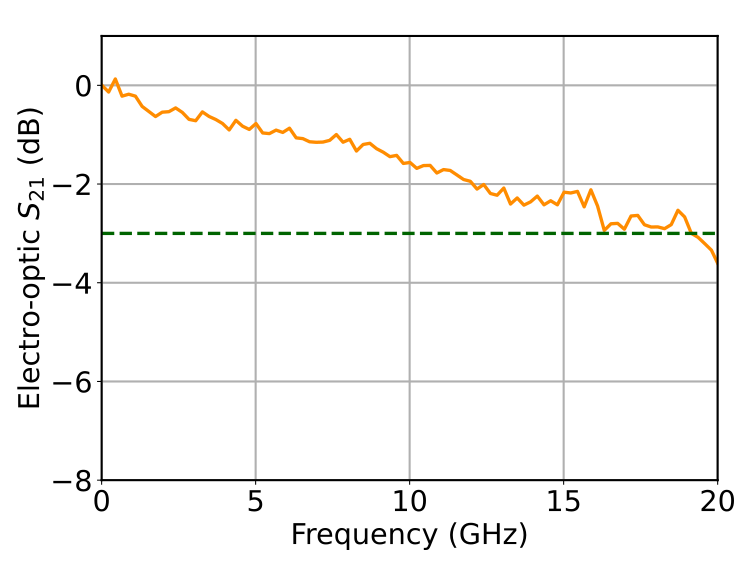}
    \caption{Bandwidth of the modulator. The figure illustrates the frequency response of the modulator, demonstrating E/O response up to 20 GHz.}
    \label{fig:modulator_bandwidth}
\end{figure}

Commercial LiNbO$_3$ intensity modulators are employed to encode the electrical signal into the optical carrier, we characterize the modulator bandwidth over a wide wavelength range, the measured bandwidth is about 20 GHz and shows high consistency over different wavelengths. As shown in Fig.~\ref{fig:modulator_bandwidth}.

\newpage
\subsection{Subsection 2: Static and Dynamic modulator transfer function}
\label{sec:s22}
The output electric field of a non-chirped single-drive Mach-Zehnder Modulator (MZM) could be written as:
\begin{equation}
E_{\text{out}}(t) = \gamma \, E_{\text{in}}(t) \sin\left( \frac{\pi}{2V_\pi} \big(V(t) + V_{\text{bias}}\big) \right)
\label{eq:efield_transfer}
\end{equation}
The corresponding optical output power is:
\begin{equation}
P_{\text{out}}(t) = \gamma^2 P_{\text{in}} \frac{1 - \cos\left(\frac{\pi}{V_\pi} \big(V(t) + V_{\text{bias}}\big)\right)}{2}
\label{eq:power_transfer}
\end{equation}
where:
\begin{center}
\begin{tabular}{@{}p{0.4\textwidth}@{\hspace{2cm}}p{0.4\textwidth}@{}}
    \textbullet \ $E_{\text{out}}(t)$: Output electric field. & \textbullet \ $P_{\text{out}}(t)$: Optical output power. \\
    \textbullet \ $\gamma$: Optical amplitude scaling factor. & \textbullet \ $E_{\text{in}}(t)$: Input electric field. \\
    \textbullet \ $P_{\text{in}}$: Input optical power ($P_{\text{in}} = |E_{\text{in}}(t)|^2$). & \textbullet \ $V(t)$: Applied voltage signal. \\
    \textbullet \ $V_{\text{bias}}$: DC bias voltage. & \textbullet \ $V_\pi$: Voltage required for a $\pi$-phase shift. \\
\end{tabular}
\end{center}

\begin{figure}[H]
    \centering
    \includegraphics[width=0.35\textwidth]{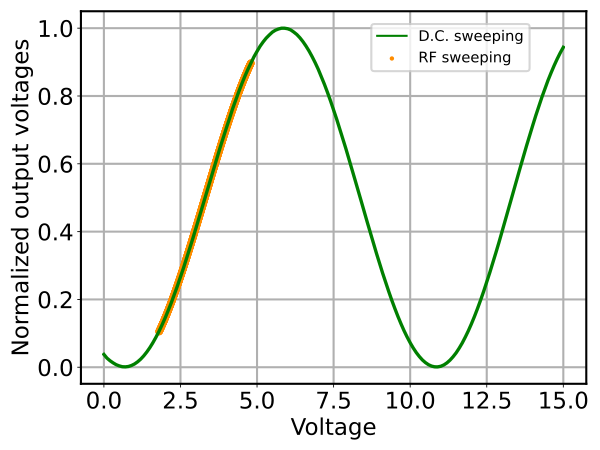}
    \caption{Power transfer function of the modulator. The green line illustrates the D.C. response of the modulator output power over a wide voltage range where the orange dot represents the response when sending random high-speed data from the AWG to the modulator and measuring with photodetector and oscilloscope, these two responses match well.}
    \label{fig:modulator_transfer_function}
\end{figure}

We modeling the modulator power transfer function using two methods, the first method is performing D.C. voltage sweep over a wide voltage range and measuring the output power using a photodetector and an oscillascope, another method is setting the D.C. bias voltage at the power transfer function quadrature point and then apply random AWG RF voltage within the AWG $V_{\text{pp}}$ to the modulator and measure the output power, after the measurement, we fit the power transfer function using  eq.~\eqref{eq:power_transfer}, the fitted power transfer function obtained using these two methods matched well, as shown in Fig.~\ref{fig:modulator_transfer_function}.

\newpage
\subsection{Subsection 3: Perform analog value multiplication using E/O modulators}
\label{sec:s23}
\begin{figure}[H]
    \centering
    \includegraphics[width=1\textwidth]{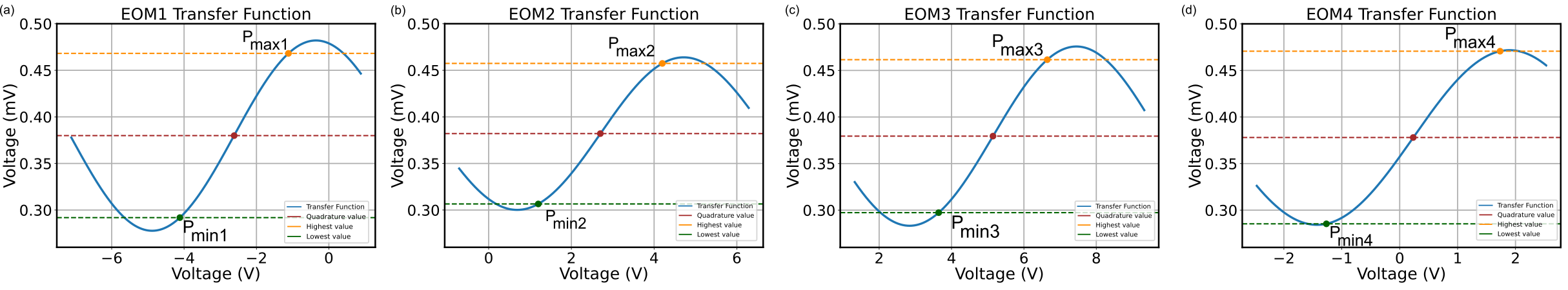}
    \caption{Power transfer function of different modulators. All these measured E/O modulator transfer functions are different and have different calibration parameters.}
    \label{fig:modulator_group_transfer_function}
\end{figure}

We use the optical hardware to perform the analog value multiplication, in order to do this, we need to map the desired analog values to the E/O modulator output powers, and extract the corresponding modulator input voltages by solving each modulator's nonlinear transfer function individually, because all these 16 modulators used in the experiment have different transfer function fitting parameters, as illustrated in Fig.~\ref{fig:modulator_group_transfer_function}.
Here, we take an example of using four modulators to show how this process works. Fig.~\ref{fig:modulator_group_transfer_function} shows the transfer functions of the four modulators from one weight modulator group. The highest and lowest output powers of these four modulators are \( P_{\text{max1}} \), \( P_{\text{max2}} \), \( P_{\text{max3}} \), \( P_{\text{max4}} \), and \( P_{\text{min1}} \), \( P_{\text{min2}} \), \( P_{\text{min3}} \), \( P_{\text{min4}} \), respectively. We need to make sure for all these modulators, the same floating numbers correspond to the same modulator output power values, so we select the highest value from \( P_{\text{min1}} \), \( P_{\text{min2}} \), \( P_{\text{min3}} \), \( P_{\text{min4}} \), which is \( P_{\text{min2}} \) to map the lowest floating number \( P_{\text{min}} \), and lowest value from \( P_{\text{max1}} \), \( P_{\text{max2}} \), \( P_{\text{max3}} \), \( P_{\text{max4}} \), which is \( P_{\text{max2}} \) to map the maximum floating point value \( P_{\text{max}} \). Once we mapped the minimum and maximum analog value numbers to \(P_{\text{min}}\) and \(P_{\text{max}} \), any analog value numbers between [\(P_{\text{min}} \),\( P_{\text{max}}\)] could be mapped to the output power according to eq.~\eqref{eq:output_voltage}  
\begin{equation}
P_{\text{out}} = (Analog - Analog_{\text{min}}) *\frac{P_{\text{max}} - P_{\text{min}}}{Analog_{\text{max}} - Analog_{\text{min}}} + P_{\text{min}}
\label{eq:output_voltage}
\end{equation}
where:
\begin{center}
\begin{tabular}{@{}p{0.5\textwidth}@{\hspace{-0.5cm}}p{0.6\textwidth}@{}}
    \textbullet \ $P_{\text{max}}$: Maximum photodetector output power. & \textbullet \ $P_{\text{min}}$: Minimum photodetector output power. \\
    \textbullet \ $Analog_{\text{max}}$: Maximum analog value power.& \textbullet \ $Analog_{\text{min}}$: Minimum analog value power. \\
    \textbullet \ $Analog$: Desired analog value. & \textbullet \ $P_{\text{out}}$: Desired analog value mapped output Power. \\
\end{tabular}
\end{center}
Now we have mapped the analog values to the photodetector output voltages. To extract the corresponding AWG input voltages, we calibrate the different E/O modulator transfer functions, which are described in eq.~\eqref{eq:power_transfer}.
From eq.~\eqref{eq:power_transfer}, the extracted input voltage is 
\begin{equation}
V(t) = \frac{V_\pi}{\pi}\arccos\left(1-2\frac{P_{\text{out}}(t)}{\gamma^2 P_{\text{in}}}\right) - V_{\text{bias}}.
\label{eq:voltage_extraction}
\end{equation}
Now we can map the desired analog value into the E/O modulator input voltage, all these E/O modulators have different transfer functions and output analog values so the input voltage extraction for each modulator needs to be performed individually. After the encoding, we can perform analog value multiplication and accumulation operations by sending different AWG voltages into the modulators.
In a DNN, the output of the analog value multiplication will serve as the input for the next layer, so we also need to perform the decoding from the photodetector output power to the analog value.
For our architecture, in a single time step, 4 MACs are performed through wavelength and space multiplexing. So we use the fact that
\begin{equation}
0*0+0*0+0*0+0*0=0
\label{eq:zero_sum}
\end{equation}
\begin{equation}
1*1+1*1+1*1+1*1=4
\label{eq:one_sum}
\end{equation}
According to eq.~\eqref{eq:zero_sum} and eq.~\eqref{eq:one_sum}, we set all the E/O modulator output power to the largest analog value and the photodetector output power \( P_{\text{maxpd}}\), which is the maximum case, corresponds to the analog value \( 4*Analog_{\text{max}}\); when setting all the E/O modulator output power to smallest analog value, the photodetector has the lowest output power \( P_{\text{minpd}}\) which corresponds to analog value \( 4*Analog_{\text{min}}\).
This principle could also be used to calibrate the analog time integration, for the time integrator that integrates over \( l\) time steps, the maximum photodetector output power corresponds to the analog value \( 4*l*Analog_{\text{max}}\) and the minimum output power corresponds to the analog value \( 4*l*Analog_{\text{min}}\). Any output powers between the maximum output power and the minimum output power could be mapped to their corresponding analog value through 
\begin{equation}
Analog_{\text{out}} = (P_{\text{out}}- P_{\text{outmin}}) *\frac{Analog_{\text{max}} - Analog_{\text{min}}}{P_{\text{outmax}} - P_{\text{outmin}}} + Analog_{\text{min}}
\label{eq:decoding floating}
\end{equation}
Using these above-mentioned encoding and decoding methods, we perform the Analog value multiplication using the optical hardware, and the optical computing experimental results match well with the theory value, showing high computing accuracy over 8 bits.

\subsection{Subsection 4: Scaling to high sampling rate}
\label{sec:s24}
\begin{figure}[H]
    \centering
    \includegraphics[width=0.8\textwidth]{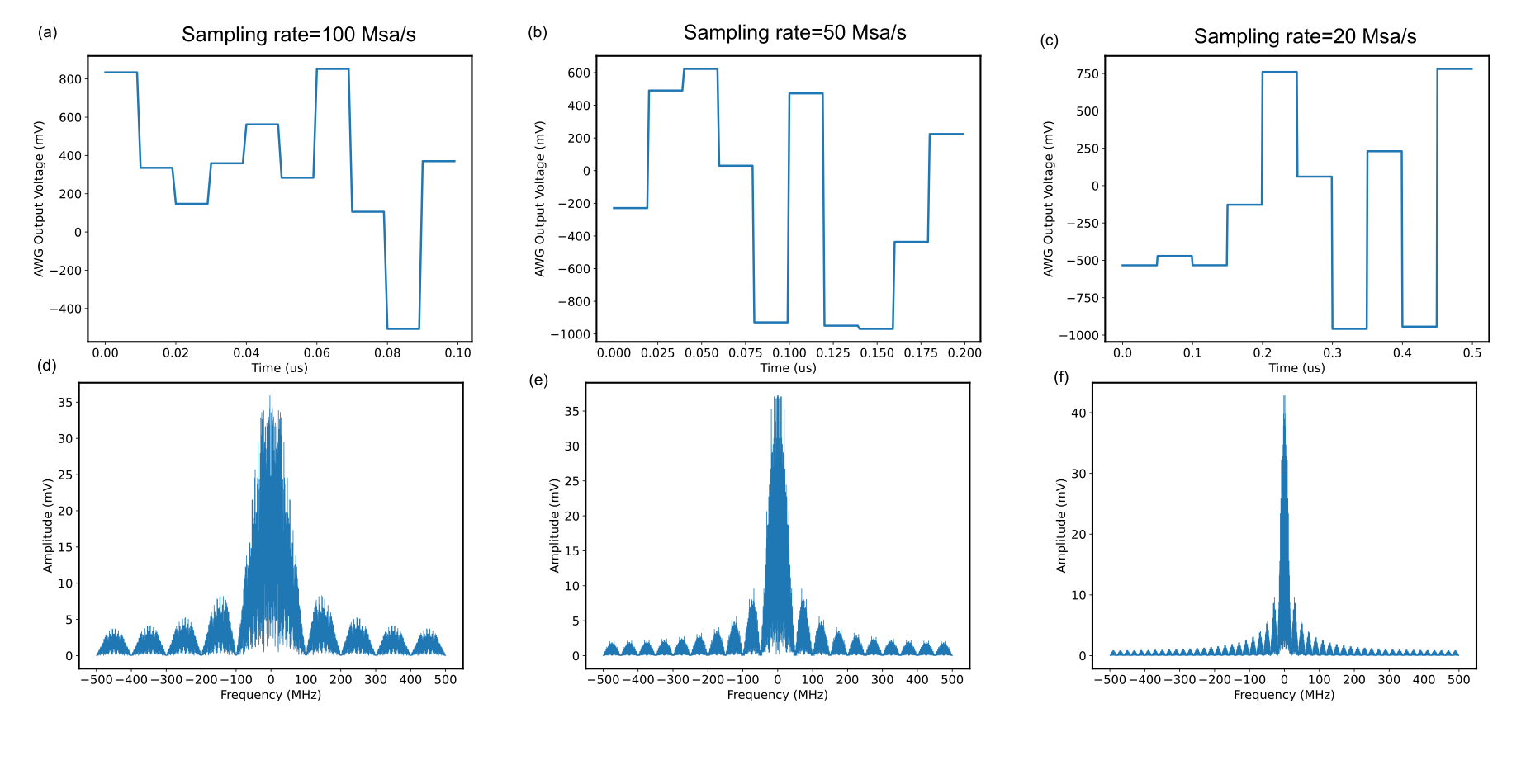}
    \caption{Perform pulse amplitude modulation of the AWG pattern, the top figure shows the pattern after the pulse amplitude modulation and the bottom figure shows the frequency information of these signals, high sampling rate modulation includes more high-frequency component data.}
    \label{fig:pulse_amplitude_modulation}
\end{figure}

\begin{figure}[H]
    \centering
    \includegraphics[width=0.38\textwidth]{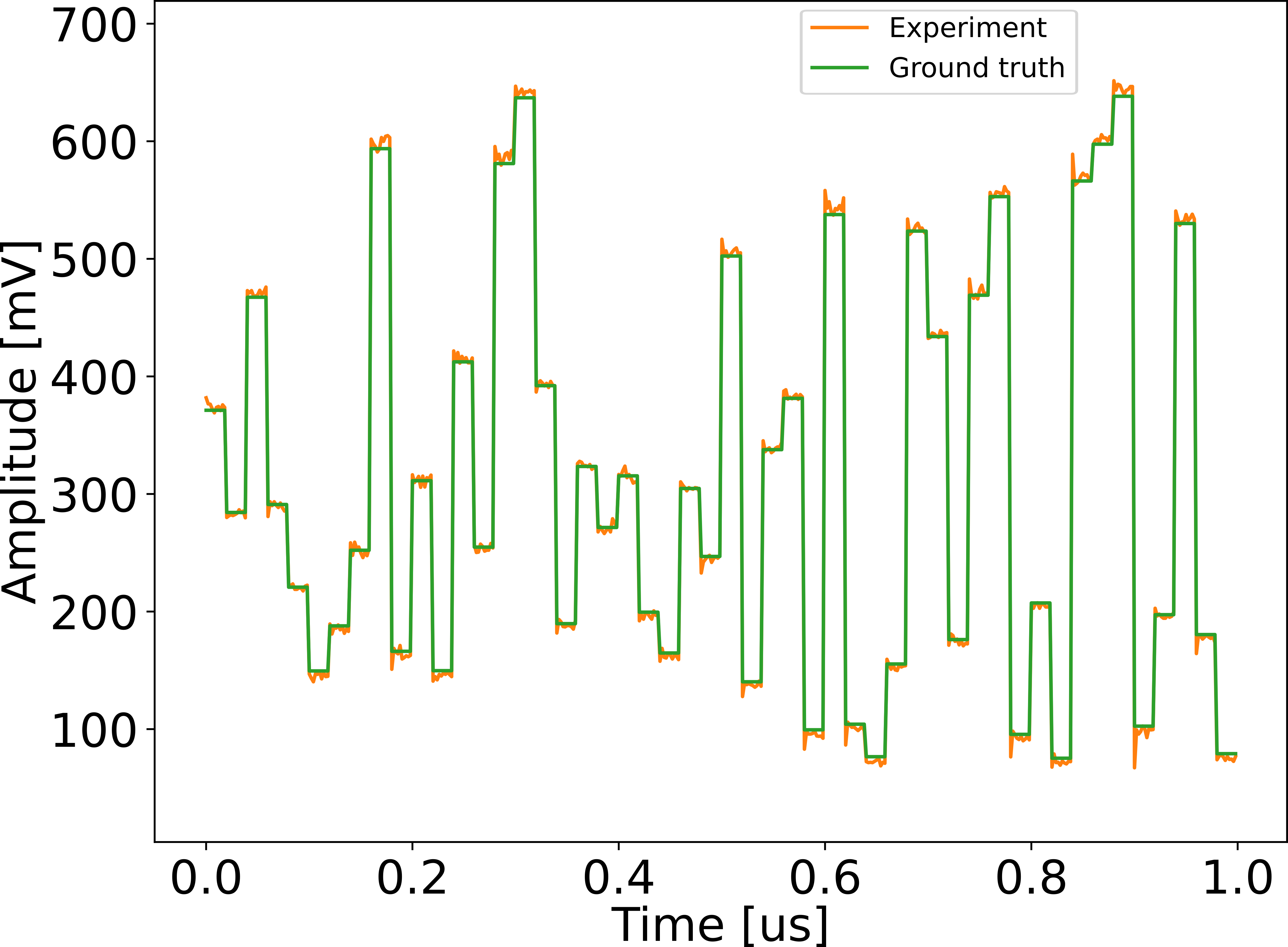}
    \caption{Pulse train of the calculated and measured signal at \( 50\,\text{MSa}\,\text{s}^{-1} \). }
    \label{fig:theory_experiment_train}
\end{figure}

\begin{figure}[H]
    \centering
    \includegraphics[width=0.38\textwidth]{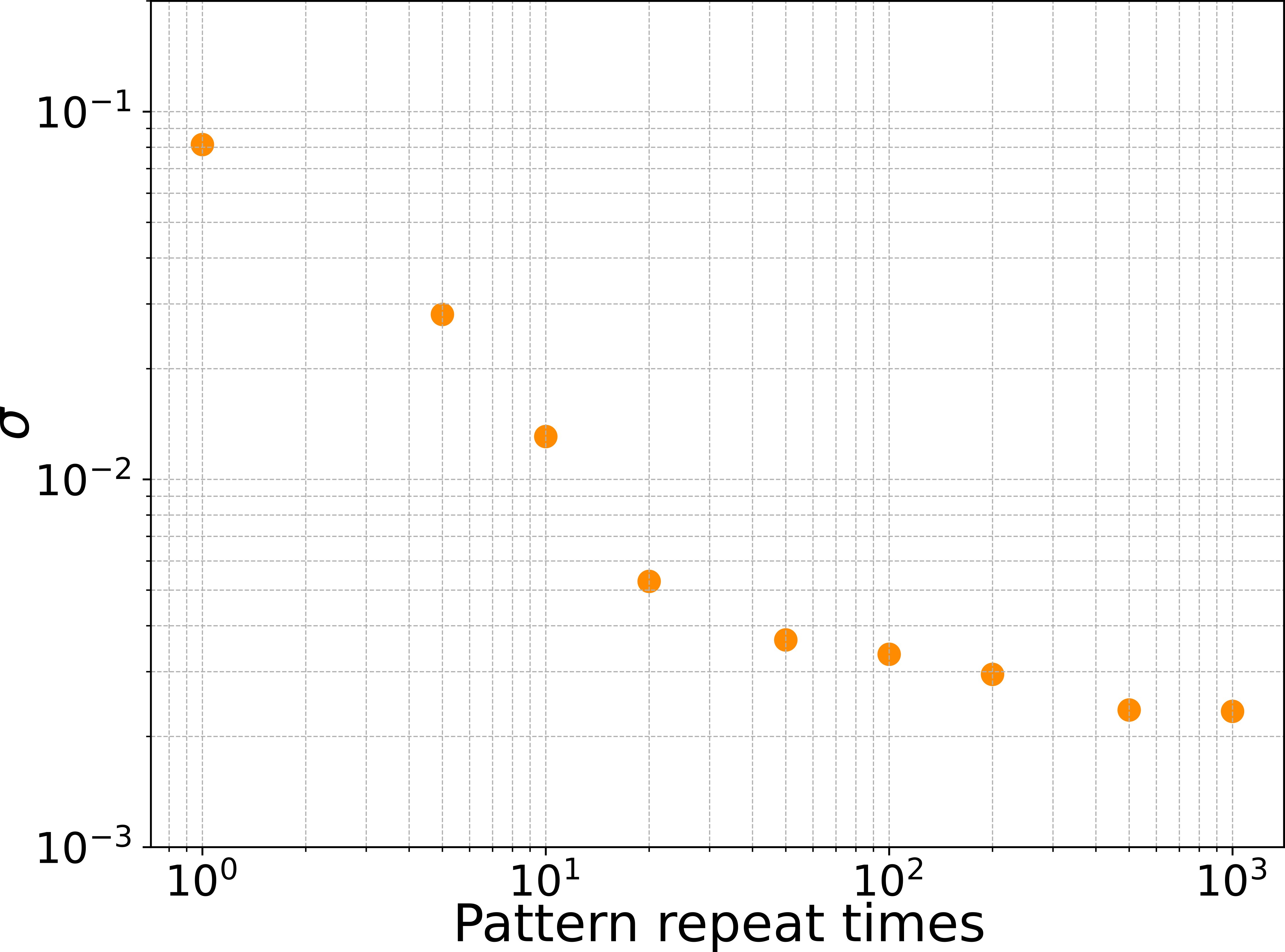}
    \caption{Experiment-theory error standard-deviation at different data sampling rates.}
    \label{fig:std_dev_versus_samplingrate}
\end{figure}

We investigate the optical architecture computing bit precision under different data encoding speeds. The AWG we used supports multiple channel implementation, the highest sampling rate is \( 1\,\text{GSa}\,\text{s}^{-1} \), and the highest analog bandwidth is 400 MHz. In order to keep a high bit precision, we down-sample the AWG encoding speed using pulse amplitude modulation. Fig.~\ref{fig:pulse_amplitude_modulation} shows AWG patterns with different sampling rates in the time and frequency domain, different sampling rate AWG signals have different frequency domain waveforms and components, we set the sampling rate to \( 50\,\text{MSa}\,\text{s}^{-1} \), which is limited by the bandwidth (65 MHz) of the ADC used in the experiment. Further increase the sampling rate will lose the high-frequency components of the AWG pattern and cause reduced computing accuracy.
Fig.~\ref{fig:theory_experiment_train} shows the pulse train of the calculated, measured signal traces and the ground truth at \( 50\,\text{MSa}\,\text{s}^{-1} \) sampling rate. The experimental trace and theory trace match well, we extract the central value from the measured pattern at each value level and compare with the ground truth, the calculated error standard deviation is 0.003, which corresponds to over 8 bits computing precision. Fig.~\ref{fig:std_dev_versus_samplingrate} shows the error at different sampling rates. Based on the trade-off of the bandwidth and accuracy, the sampling rate is kept at \( 50\,\text{MSa}\,\text{s}^{-1} \).

\subsection{Subsection 5: Stabilizing quadrature point drift }
\label{sec:s25}
\begin{figure}[H]
    \centering
    \includegraphics[width=1\textwidth]{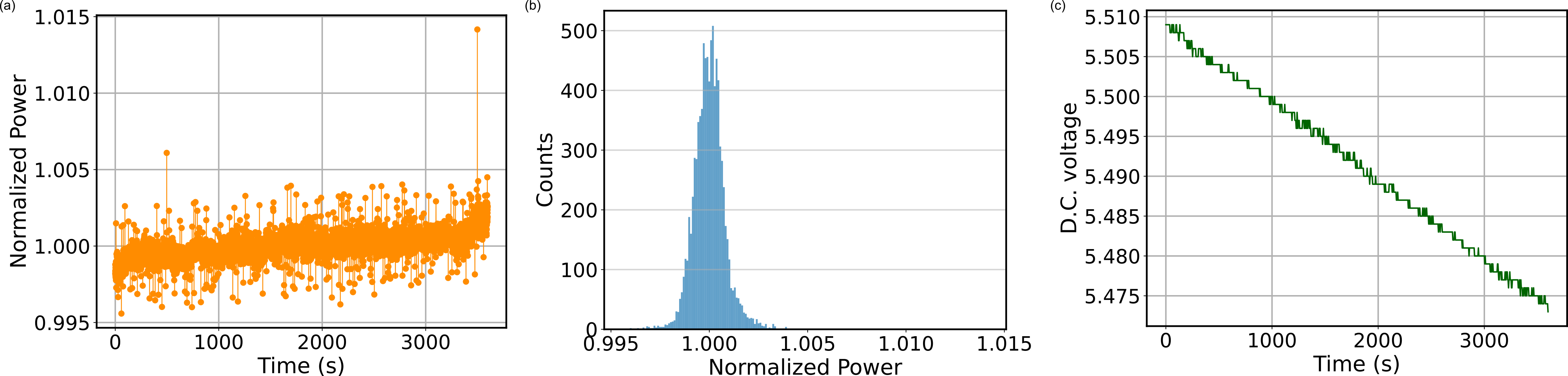}
    \caption{E/O modulator quadrature-point drift measurements. The output power is very stable when the D.C. bias voltage stabilization system is employed.}
    \label{fig:quadrature-point drift}
\end{figure}

LiNbO$_3$ modulator is known to be susceptible to quadrature-point drift, where the voltage corresponding to a fixed output power varies over time, which is attributed to fluctuations in charges accumulated in the phase shifters, this quadrature-point drift happens on the time scale of several seconds. We use a photodetector to provide feedback to D.C. voltage to stabilize the modulator quadrature-point output power. Fig.~\ref{fig:quadrature-point drift} shows the E/O modulator outputs when the D.C. bias stabilization is employed, the D.C. voltage is gradually decreased, which is used to compensate for the quadrature-point shift and the output power is very stable, thanks to the stable modulator output power and constant transfer function, the modulator calibration just needs to be performed once and the system shows high bit precision. 

\newpage
\section{Free space grating beam routing system as 3D parallel wavelength Mux and DeMux}
The \( 4*4\) grating beam routing system works as parallel, 64-channel input, and 64-channel output Mux and DeMux. We characterize the system performance when the grating beam routing system works as a Mux and DeMux, respectively. 

\subsection{Subsection 1: Grating beam routing system parallelism characterization }
\label{sec:s31}
\begin{figure}[H]
    \centering
    \includegraphics[width=0.7\textwidth]{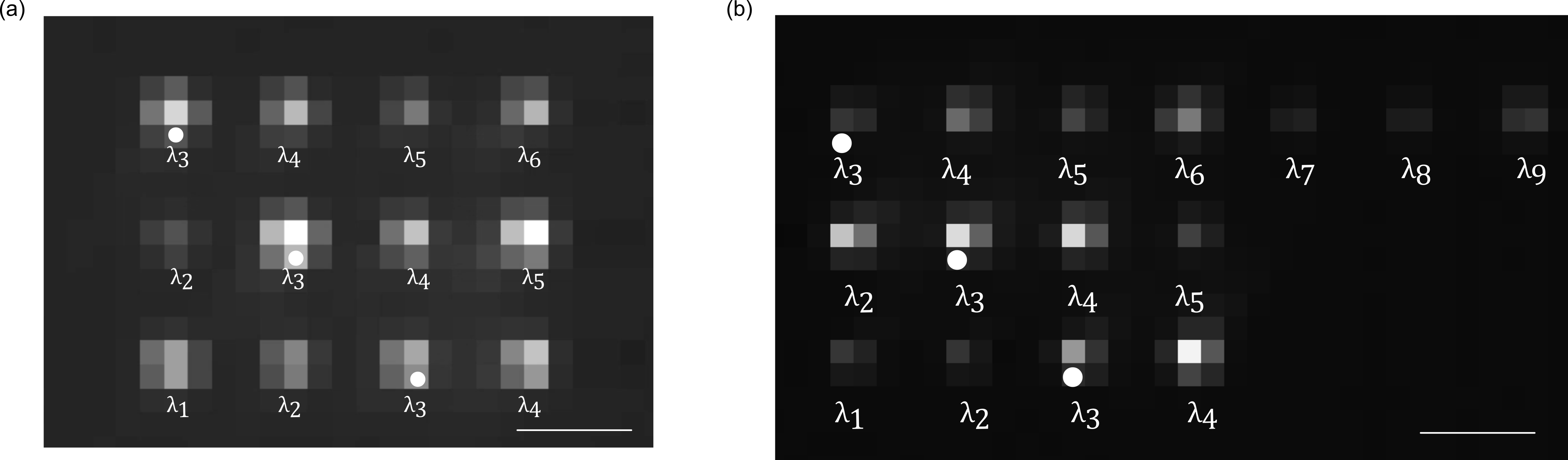}
    \caption{Parallelism measurement of the grating beam routing architecture, the white dot represents the light incident spatial channel and the wavelengths indicate the frequency of the output light that diffracted to this spatial channel.}
    \label{fig:grating_characterization}
\end{figure}

\begin{figure}[H]
    \centering
    \includegraphics[width=0.3\textwidth]{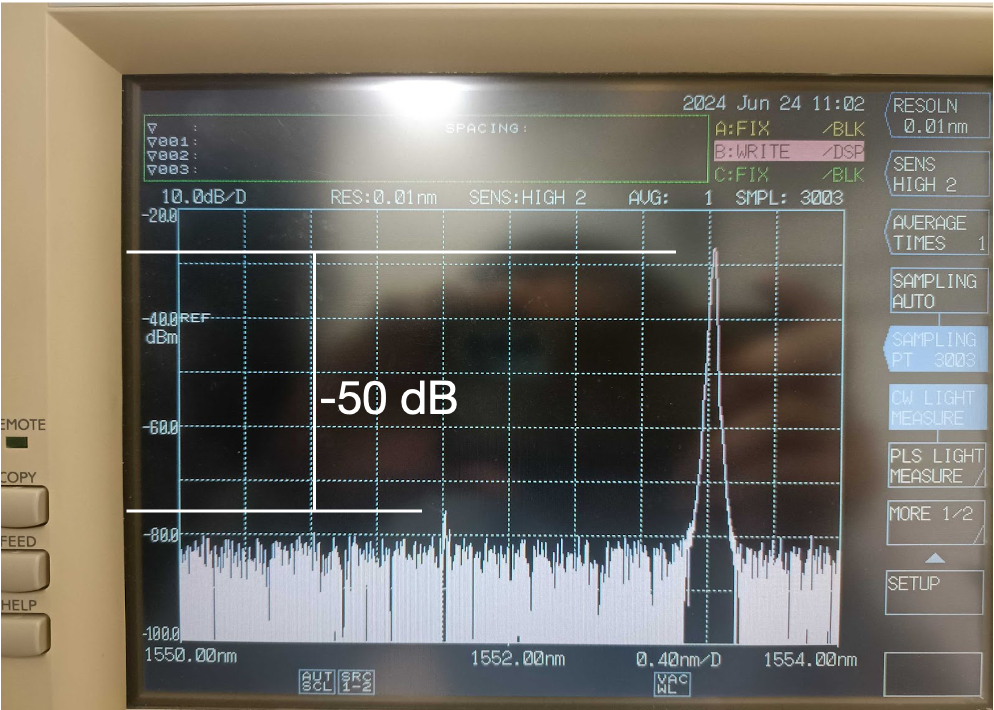}
    \caption{Crosstalk measurement of the grating beam routing system, the crosstalk between different channels are -50 dB.}
    \label{fig:crosstalk}
\end{figure}

We characterize the hardware parallelism when sending multiple wavelengths from different channels. To make the results more evident, we let the light incident from different rows of the fiber array so the output distribution can easily be separated.
Fig.~\ref{fig:grating_characterization} (a) shows the output image from the receiver fiber array when we send four different wavelengths into each channel, the wavelengths in each different channel are offsetetd with one WDM spacing.  The output wavelengths are equally distributed among different channels, adjacent wavelength lights incident from adjacent spatial channels will emitted into the same column of the receiver fiber array. We further investigate the broadband operation of the grating architecture where seven different wavelengths are injected to the grating through the same channel and the output lights are equally distributed to seven different positions, these spots have different intensity because the input light intensity for different wavelengths are different.
The grating architecture also has low crosstalk, we measure the crosstalk between different channels, and the measured result is about -50 dB, as indicated in Fig.~\ref{fig:crosstalk}

Having demonstrated the high parallelism and low crosstalk advantages of the grating beam routing system, we characterize the performance of the grating when working in the optical computing architecture as a wavelength Mux and DeMux.

\subsection{Subsection 2: Grating beam routing system works as Mux}
\label{sec:s32}

\begin{figure}[H]
    \centering
    \includegraphics[width=0.68\textwidth]{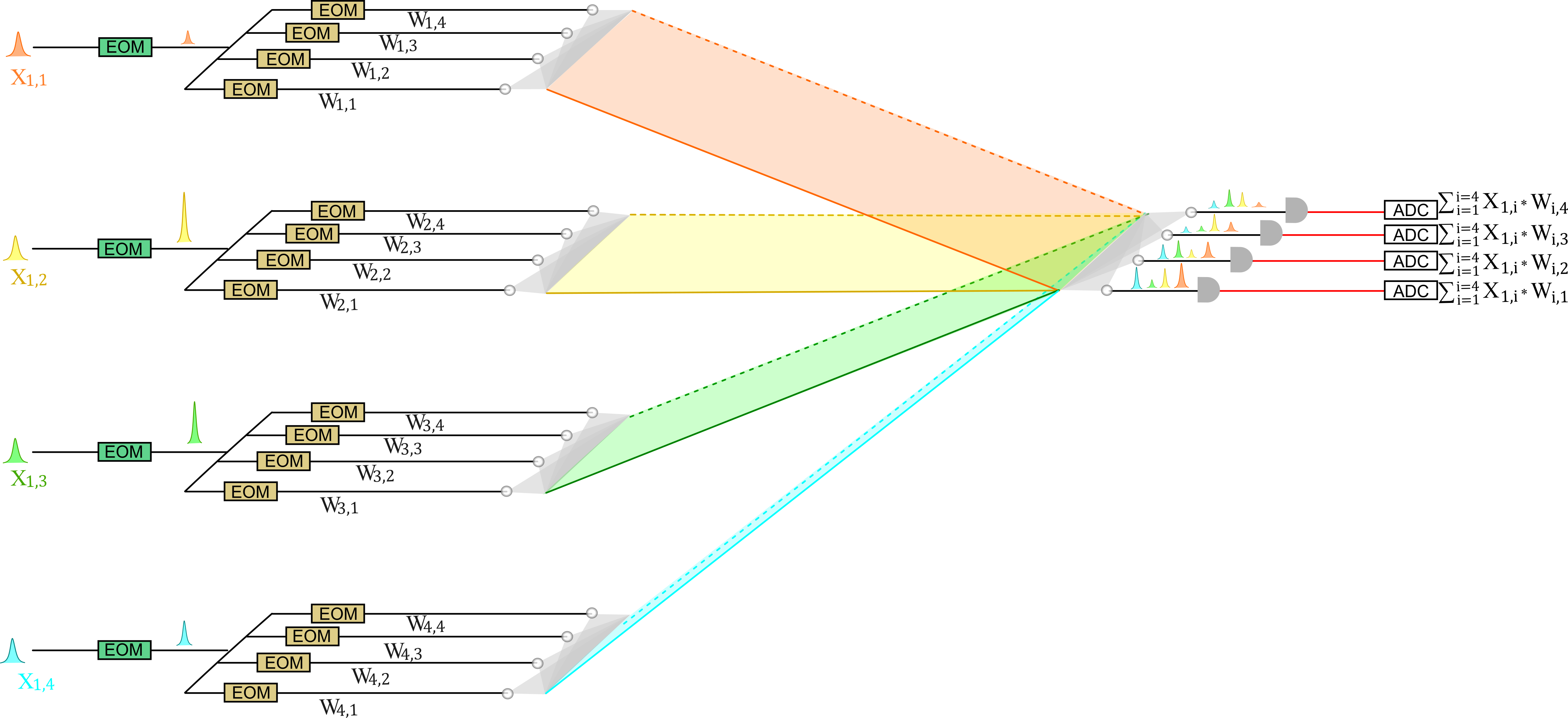}
    \caption{Architecture setup when the grating beam routing system works as Mux.}
    \label{fig:mux_setup}
\end{figure} 
\begin{figure}[H]
    \centering
    \includegraphics[width=0.68\textwidth]{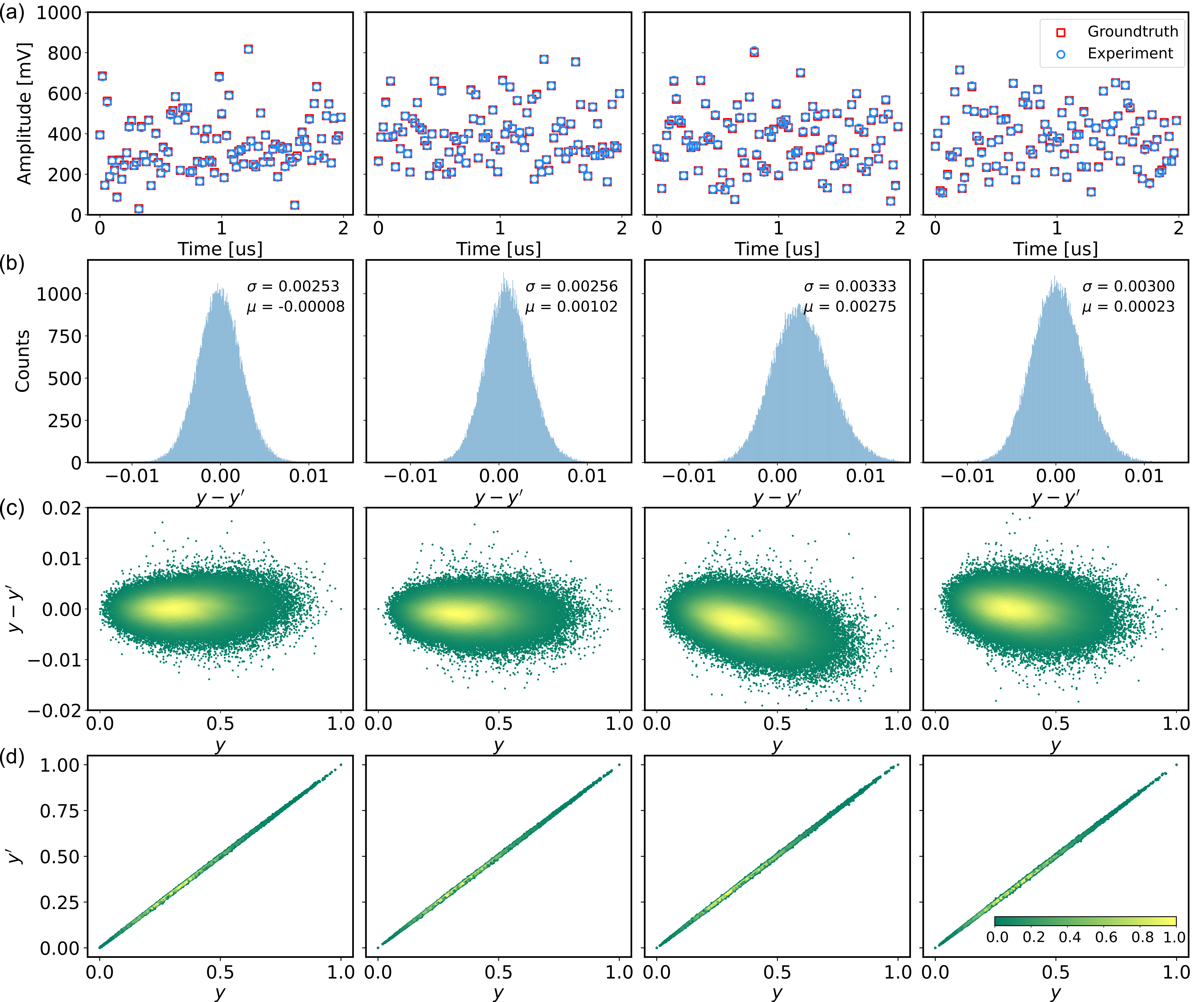}
    \caption{Experimental results of the grating beam routing Mux.}
    \label{fig:mux_results}
\end{figure} 
Fig.~\ref{fig:mux_setup} shows the setup of the system when employing the grating beam routing system as wavelength Mux. Four data modulators are modulated by four different wavelengths, each modulated signal is then spatially fan-out to four weight modulators, the outputs are then guided to the grating beam routing system through fiber array, the grating beam routing system works as wavelength/spatial Mux, different wavelengths are multiplexed to the same channel and the outputs are being recorded by amplified photodetectors.
Fig.~\ref{fig:mux_results} shows the measured waveform and MAC error distribution. The modulation signal sampling rate is \( 50\,\text{MSa}\,\text{s}^{-1} \). The system shows low error standard deviation and high computing accuracy across multiple channels simultaneously.

\subsection{Subsection 3: Grating beam routing system works as DeMux}
\label{sec:s33}

\begin{figure}[H]
    \centering
    \includegraphics[width=0.68\textwidth]{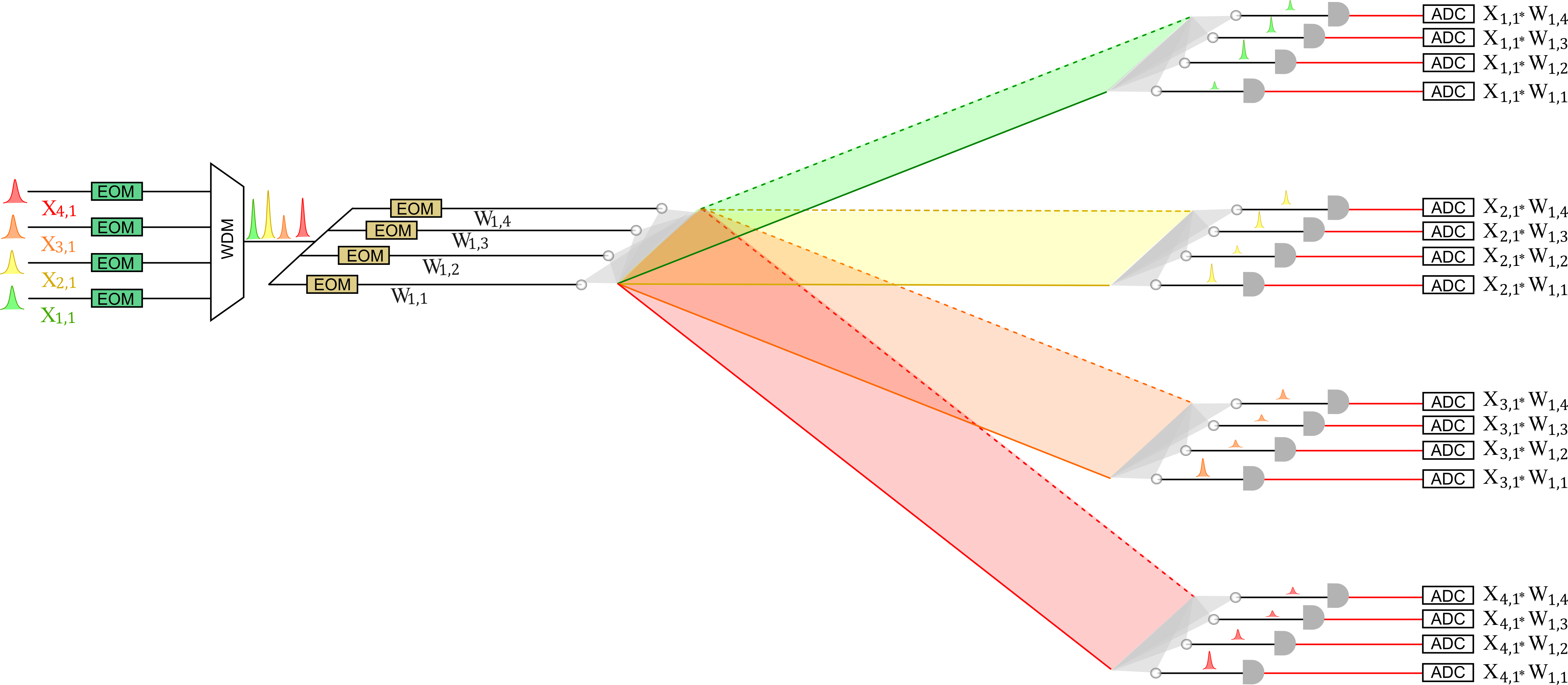}
    \caption{Architecture setup when the grating beam routing system works as DeMux. }
    \label{fig:demux_setup}
\end{figure} 
\begin{figure}[H]
    \centering
    \includegraphics[width=0.68\textwidth]{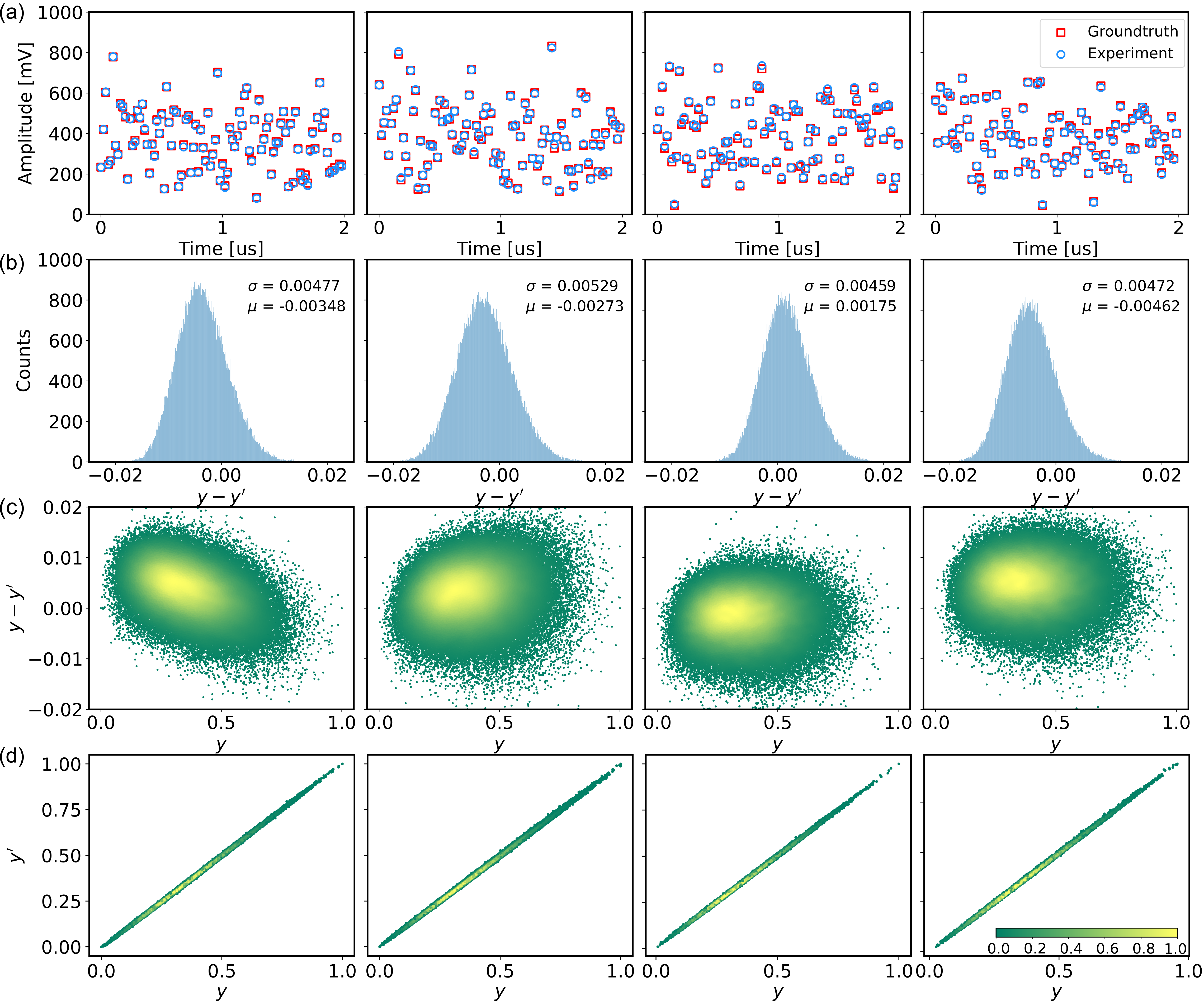}
    \caption{Experimental results of the grating beam routing DeMux. }
    \label{fig:demux_results}
\end{figure} 
The grating beam routing system also works as wavelength DeMux. Fig.~\ref{fig:demux_setup} shows the setup of the system when a commercial WDM multiplex four different wavelengths from different channels into one channel and then spatially fan-out to four different weight modulators, the modulated signals are then demultiplexed by the grating, the system structure is similar to our previous work \cite{sludds_netcast} but provides 3D data parallelism advantage.  
Fig.~\ref{fig:demux_results} shows the measured waveform and multiplication error distribution. The system shows low error standard deviation and high computing accuracy, the results are comparable with work \cite{sludds_netcast}, which uses commercial WDM for DeMux and has the system parallelism of $O(N)$, while our grating architecture provides $O(N^3)$ parallelism with only one device.
\subsection{Subsection 4: Minimum wavelength spacing}
\label{sec:s34}
\begin{figure}[H]
    \centering
    \includegraphics[width=0.3\textwidth]{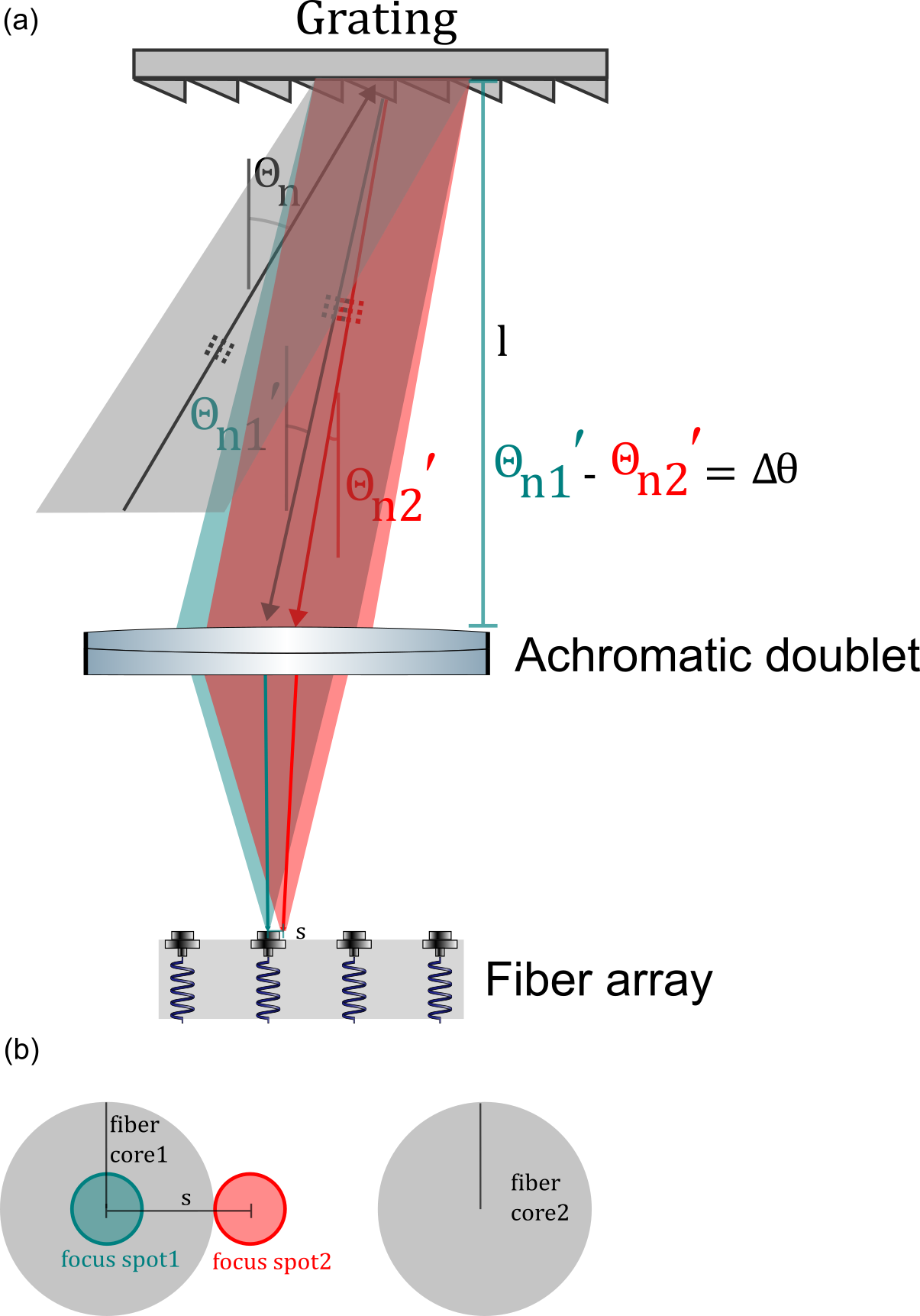}
    \caption{Concept of how to calculate the minimum wavelength spacing of the grating beam routing architecture.}
    \label{fig:minimum wavelength}
\end{figure}

The minimum wavelength spacing is determined by the crosstalk of these two wavelengths beams when these two beams are focused by the lens in the surface of the receiver fiber array.
Fig.~\ref{fig:minimum wavelength} shows the concept of how the minimum wavelength spacing is calculated. Broadband light emitted from the input fiber array becomes collimated after the input achromatic doublet, the beam diameter, which is proportional to the focal lens, is about 1.9 \,\text{cm}. Through the grating diffraction, different wavelength components will be scattered at different angles, with the angle difference $\Delta\theta$ determined by the grating angular dispersion, and the distance between these two diffracted beams are determined by both the grating angular dispersion and the distance \( l\) between the grating and the receiver achromatic doublet. After propagate the distance \( l\), these two collimated beams are focused in the front surface of the receiver fiber array by the receiver achromatic doublets.
The minimum wavelength spacing without channel/wavelength crosstalk is shown in the Fig.~\ref{fig:minimum wavelength}. In this figure, the shorter wavelength is focused in the center of the receiver fiber array, where another wavelength beam is just focused outside the receiver fiber array core, in this case, the receiver fiber array will only collect the lights from the shorter wavelength.
The minimum distance between the focused spots is determined by the receiver fiber array core diameter, which is 50\,\textmu m. The numerical aperture (NA) of the multi-mode fiber is 0.22. This corresponds to a critical angle of \ang{12.7}, which is larger than the half solid angle \ang{7.2} of the focused beam. The beam waist is 8\,\textmu m, so the focused beam could be totally collected. We also need to consider the evanescent field of the light, which could otherwise cause crosstalk.
The evanescent field penetration depth could be calculated through
\begin{equation}
d_p = \frac{\lambda}{2 \pi n_1 \sqrt{\sin^2\theta - \left(\frac{n_2}{n_1}\right)^2}}
\label{eq:peneration_depth}
\end{equation}
From Eq.~\eqref{eq:peneration_depth}, the penetration depth is calculated to be $300\,\text{nm}$. Therefore, the minimum distance \(g\) shown in the figure is 33\,\textmu m, and the diffraction angle difference \(\Delta\theta\) is given by:
\[
\Delta\theta = \frac{\SI{33}{\micro\meter}}{l}.
\]
And the corresponding wavelength difference is \begin{equation}
\Delta\lambda = \frac{1}{2} \lambda_0 \Delta\theta \cot(\theta_0),
\label{eq:delta_lambda}
\end{equation}
where:
\begin{itemize}
    \item $\theta_0$ is directly back scattered by the grating.
\end{itemize}

Consider the grating-receiver achromatic doublet distance \(l\) of 30\,\text{cm}, the wavelength difference is 
\[
\frac{\lambda d}{2l} = 0.09 \,\text{nm}.
\]
The theoretical resolution of the grating is \( \frac{\lambda}{N} = 0.057 \, \text{nm} \), which meets the low crosstalk requirement of the system.
We need to noticed that the calculated wavelength spacing is the theoretical value, in the real case, we also need to consider the scattering in the fiber surface, consider the availability of the commercial WDM that we need to use to multiplex different lasers, the wavelength spacing of 0.4 $\mathrm{nm}$ (50 GHz) is a reasonable value.

\newpage
\subsection{Subsection 5: Maximum wavelength range we can use}
\label{sec:s35}
Several factors will influence the maximum wavelength range we can use, like the total bandwidth of the laser source, the bandwidth of the commercial WDM. Here, we consider the grating beam routing architecture itself, and the main limitation for the working wavelength range is the chromatic aberration of the lens, different wavelength lights will have different focal lens, making different spot size of the focused beam in the receiver fiber array front surface.
\begin{figure}[H]
    \centering
    \includegraphics[width=0.35\textwidth]{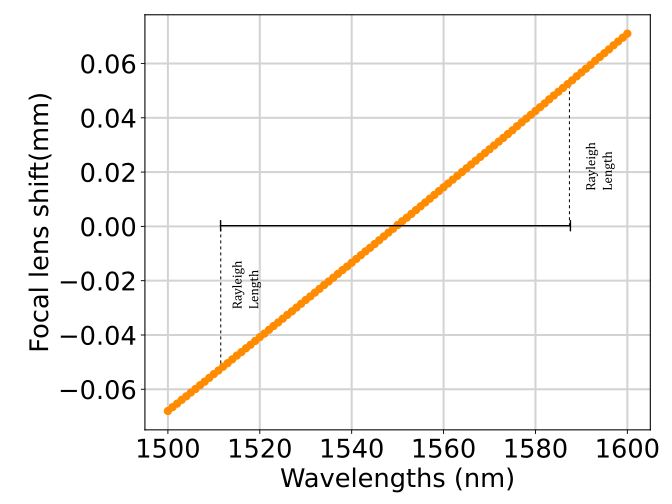}
    \caption{Focal lens shift of different wavelength lights.}
    \label{fig:focal lens shift}
\end{figure} 
\begin{figure}[H]
    \centering
    \includegraphics[width=0.35\textwidth]{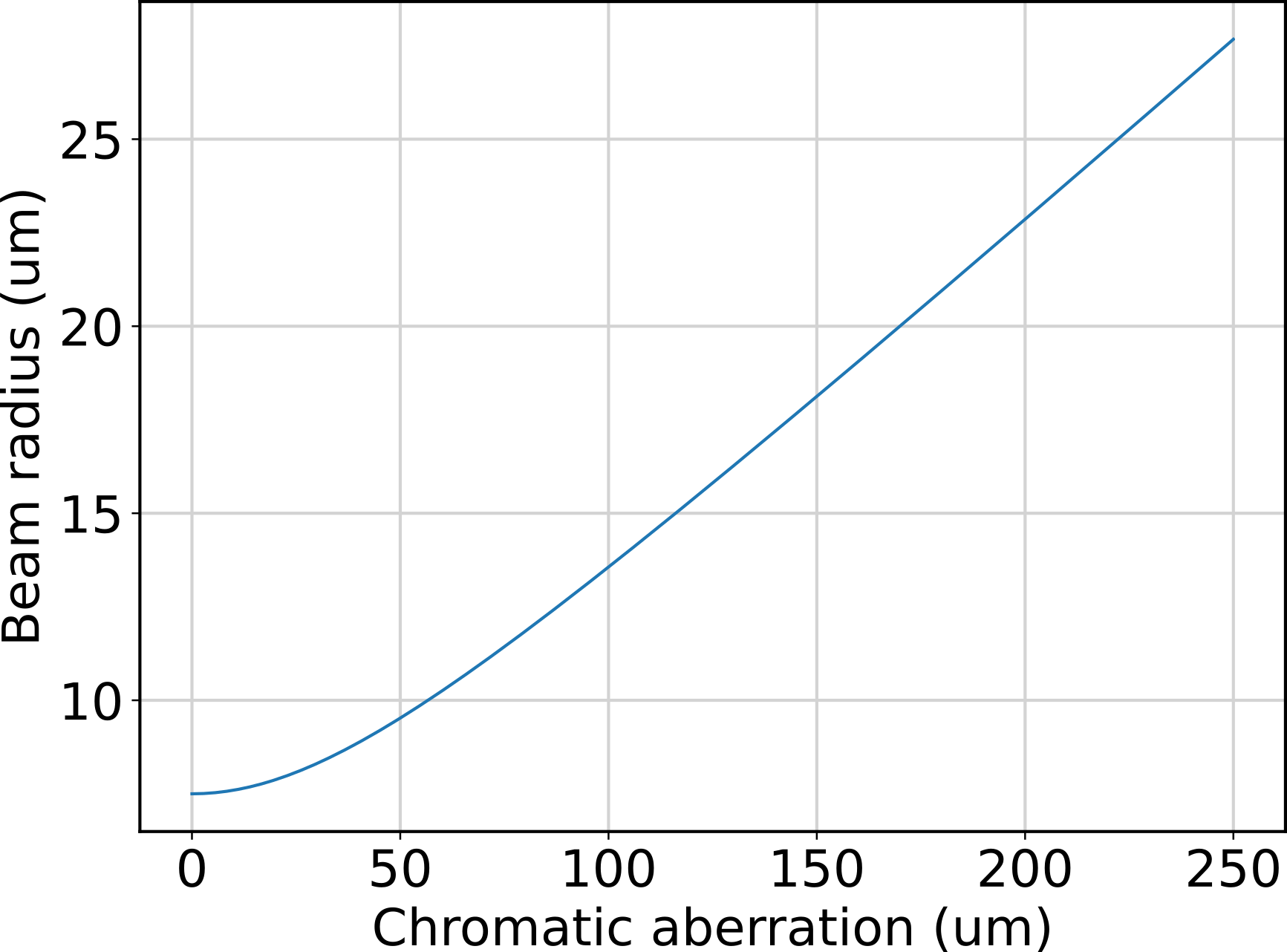}
    \caption{Spot size change caused by the chromatic aberration.}
    \label{fig:spot size drift}
\end{figure} 

Figure~\ref{fig:focal lens shift} shows the focal lens shift of the achromatic doublets. When the wavelengths change from 1530\,\text{nm} to 1580\,\text{nm}, the focal lens shift is only 70\,\textmu m.
The Rayleigh length of the focused beam is 50\,\textmu m, so the focal spot size of different wavelength lights is within the Rayleigh length range and could all be collected by the receiver fiber array. From this point of consideration, the working wavelength range is at least 50\,\text{nm}.
The large numerical aperture (NA) and core diameter of the multimode fiber also provide a large tolerance for the broadband light's chromatic aberration. Figure~\ref{fig:spot size drift} shows the spot size change versus chromatic aberration. The spot size changes from 7.5\,\textmu m to 25\,\textmu m, with a small divergence angle of \ang{7}.
So the working wavelength range of the grating beam routing architecture is over 50\,\text{nm}. The input light wavelength spacing could reach 50~GHz meanwhile keeping low crosstalk between adjacent fiber array channels, and the maximum allowed working wavelength range IS over 50 $\mathrm{nm}$, this corresponds to a wavelength channel number \(2N-1\) of 125 and fiber array channel number \(N\) over 60, which makes an off-axis distance of 4.5 mm. Achromatic doublet lenses have a much reduced sensitivity to centration on the beam axis when compared to spherical singlets and aspheric lenses, the calculated lateral and transverse aberrations is 32\,\textmu m, within the tolerance of the Rayleigh range, so the maximum working wavelength range could reach 50 $\mathrm{nm}$.

\newpage
\subsection{Subsection 6: Compare with commercial Wavelength Mux and DeMux}
\label{sec:s36}
\begin{figure}[H]
    \centering
    \includegraphics[width=0.5\textwidth]{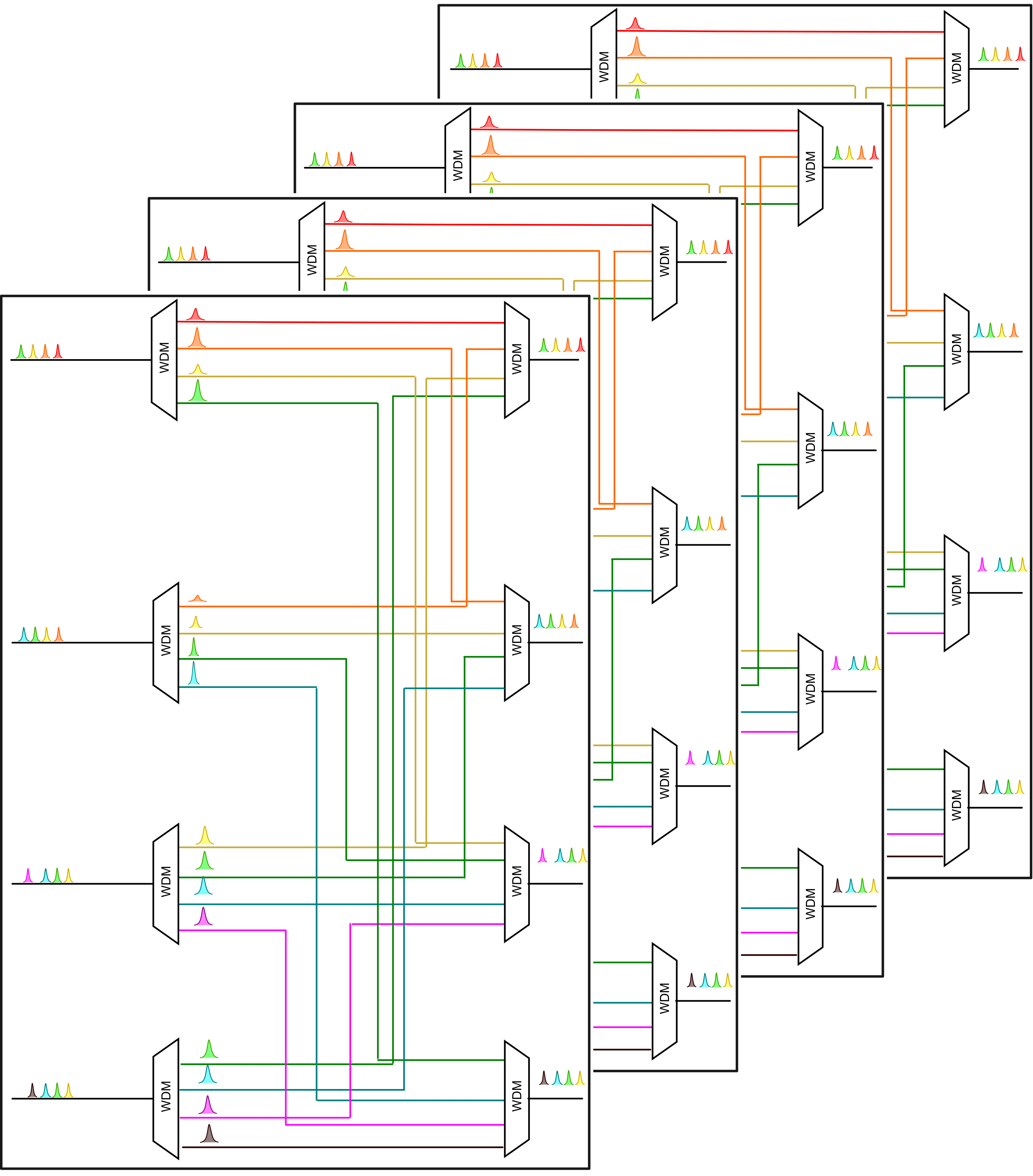}
    \caption{Schematic setup of the grating beam routing system using commercial WDM.}
    \label{fig:using commercial WDM}
\end{figure}

The current wavelength spacing of our grating beam routing architecture is 200 GHz, which is limited by the available commercial WDM we have that used to multiplex different laser wavelengths. As we discussed in the previous section, the theoretical wavelength spacing of the grating architecture is below 50 GHz, which is comparable with dense WDM, the working wavelength range is over 50\,\text{nm}, and the spatial channel could reach 120.
We characterize the commercial WDM that we used to multiplex the different wavelength lasers, the wavelength spacing for the commercial WDM is 200 GHz, with the insertion loss of 0.7 dB for the first DeMux channel, and gradually increased to 1.6 dB for the fourth channel. The insertion loss will be further increased if more WDM channels are being used.
For our grating beam router, the loss is between 0.7 dB to 1.5 dB, which is comparable with the commercial WDM. In addition, our grating beam router supports multichannel, three-dimensional input and output, which is not possible with just one commercial WDM. Let take the \( 4*4\) beam router as an example, in order to achieve the same output throughput, we need 16 commercial WDMs to DeMux the broadband wavelength into 64 different ports, and another 16 commercial WDMs to Mux the different wavelength lights from different channels into 16 Mux ports, as shown in Fig.~\ref{fig:using commercial WDM}.  
Our grating beam routing system is capable for further scaling with dense wavelength spacing. In order to achieve the same throughput of the grating beam routing system with wavelength spacing of 50 GHz and channel numbers of 120, 240 commercial WDMs are needed.

\subsection{Subsection 7: Chip scale architecture}
\label{sec:s37}
\begin{figure}[H]
    \centering
    \includegraphics[width=0.6\textwidth]{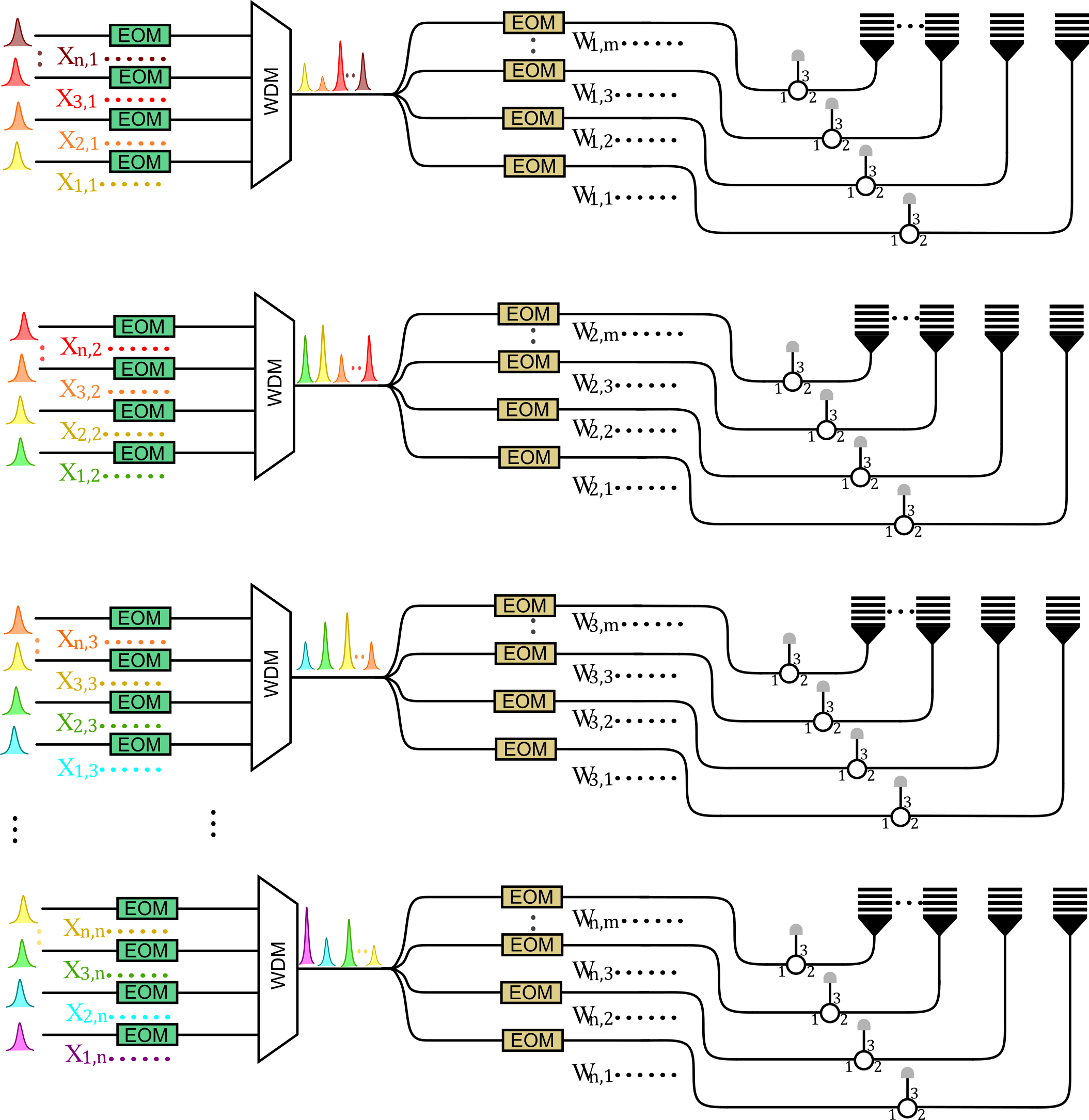}
    \caption{Concept of using chip scale modulators and fiber arrays.}
    \label{fig:chip fiber array}
\end{figure}

\begin{figure}[H]
    \centering
    \includegraphics[width=0.5\textwidth]{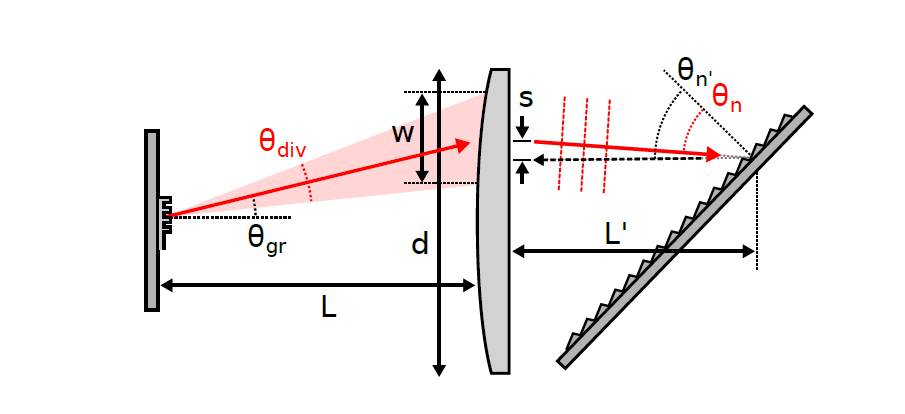}
    \caption{Effect of grating coupler angle and beam travel on the position of the return beam.}
    \label{fig:gc_diagram}
\end{figure}

The fiber array pitch could be further reduced, as illustrated in Fig.~\ref{fig:chip fiber array}. The modulators, WDMs, beam splitters, circulators and grating couplers are fabricated in the chip while the collimation lens and the blazed grating are in free space. 
The chip grating couplers are used to de-multiplex different wavelength lights into differents of the collimation lens and the free space grating according to their wavelength while the free-space grating works as Mux and DeMux by through the wavelength-spatial coupling. In this case, the chip grating pitch could be small.
The grating coupler angle does not change things as long as the emitted light is still captured by the lens. To show this, consider Fig.~\ref{fig:gc_diagram}. The grating coupler emits a cone of light at an angle $\theta_\text{gr}$ with a divergence $\theta_\text{div} = 2 \text{NA}$. Everything works in the small-NA limit, so this leads to a spot of size $w = L\theta_\text{div}$ on the lens. The lens converts this diverging spot to a collimated beam of size $w$ that propagates as a plane wave since the distance to the grating is much less than the Rayleigh length. The grating deflects the angle by an amount $|\theta_{n'} - \theta_n|$, which changes the focal point on the chip.

There is a small amount of beam travel $s = \frac{L'}{|\theta_{n'}-\theta_n|}$ due to the angular change, which leads to a beam angle deflection $\Delta\theta_\text{gr}=(s/w)\theta_\text{div}$ when focused to the PIC. As long as $s\ll w$, the beam angle is deflected by much less than the grating coupler divergence, so the light couples near-perfectly into the grating coupler. We find that (for a $45^\circ$ grating tilt):
\begin{equation}
\frac{s}{w}=\frac{L'|\theta_{n'}-\theta_n|}{w}\leq\frac{d}{w}|\theta_{n'}-\theta_n|\leq\frac{d}{w}N\Delta\theta=\frac{2d}{w}\frac{\Delta\lambda_\text{BW}}{\lambda}
\label{eq:grating_formula}
\end{equation}

\text{Equation \eqref{eq:grating_formula}} shows the relationship between the beam deflection, grating properties, and wavelength.
As long as (1) the optical bandwidth is small enough ($\Delta \lambda_\text{BW} / \lambda \ll 1$) and (2) the lens is not much bigger than the beam divergence ($d / w = O(1)$), the ratio in Eq.~\eqref{eq:grating_formula} is small and the system does not depend on the grating coupler angle $\theta_\text{gr}$ at all: light emitted at any angle will always return to the target grating coupler (with the same angle). Condition (1) is satisfied for telecom optics ($\Delta \lambda_\text{BW} / \lambda = 0.02$ for $30$ nm bandwidth). Condition (2) is related to the grating coupler angular dispersion, since the lens must be large enough to collect the light from all colors. This sets a lower bound on $d / w$:
\begin{equation}
\frac{d}{w} \geq \frac{(\theta_\text{div} + 2\Delta\theta_\text{disp})L}{\theta_\text{div}L}=\frac{\theta_\text{div} + n_g \Delta\lambda_\text{BW}/\lambda}{\theta_\text{div}}
\label{eq:lower_bound}
\end{equation}
(where we have used $d\theta \approx -n_g d\lambda / \lambda$ valid for a near-vertical grating coupler). Assuming some reasonable $n_g \lesssim 4$ and setting $\Delta \lambda = 30$ nm and $\theta_\text{div} = 2 \text{NA} = 0.28$, the ratio in Eq.~\eqref{eq:lower_bound} is around $1.25$, easily satisfying condition (2).
While this analysis suggests that grating angles are not particularly important, there are hardware solutions that may lead to more convenient vertical coupling. Micro-lens arrays or 3D-printed optics can verticalize a beam~\cite{dietrich2018situ}, at least for a target wavelength. Some SiPh processes also incorporate broadband TIR mirrors for vertical coupling~\cite{aalto2019open}.

\newpage
\section{Free space grating beam routing for parallel convolution operation}
\subsection{Subsection 1: Positive and Negative kernels }
\label{sec:s41}
For the image convolution, the image matrix data is positive and the kernel matrix contains both positive and negative numbers, we process the positive and negative kernel numbers in different time rounds.
We consider the \( 2*2\) kernel K, where a and b are positive numbers, c and d are negative numbers\[\mathbf{K} = \begin{bmatrix} a & b \\ c & d \end{bmatrix}\] The \( 28*28\) data matrix X is
\[\mathbf{X} =\begin{bmatrix}x_{1,1} & x_{1,2} & \cdots & x_{1,28} \\x_{2,1} & x_{2,2} & \cdots & x_{2,28} \\\vdots  & \vdots  & \ddots & \vdots  \\x_{28,1} & x_{28,2} & \cdots & x_{28,28}\end{bmatrix}\] 
\begin{align*}
\mathbf{Y} &= \mathbf{X} \ast \mathbf{K} \\
&=
\begin{bmatrix} 
a & b \\ 
c & d 
\end{bmatrix}
\ast
\begin{bmatrix}
x_{1,1} & x_{1,2} & \cdots & x_{1,28} \\
x_{2,1} & x_{2,2} & \cdots & x_{2,28} \\
\vdots  & \vdots  & \ddots & \vdots  \\
x_{28,1} & x_{28,2} & \cdots & x_{28,28}
\end{bmatrix} \\
&=
\begin{bmatrix} 
a & b \\ 
0 & 0 
\end{bmatrix}
\ast
\begin{bmatrix}
x_{1,1} & x_{1,2} & \cdots & x_{1,28} \\
x_{2,1} & x_{2,2} & \cdots & x_{2,28} \\
\vdots  & \vdots  & \ddots & \vdots  \\
x_{28,1} & x_{28,2} & \cdots & x_{28,28}
\end{bmatrix} +
\begin{bmatrix} 
0 & 0 \\ 
c & d 
\end{bmatrix}
\ast
\begin{bmatrix}
x_{1,1} & x_{1,2} & \cdots & x_{1,28} \\
x_{2,1} & x_{2,2} & \cdots & x_{2,28} \\
\vdots  & \vdots  & \ddots & \vdots  \\
x_{28,1} & x_{28,2} & \cdots & x_{28,28}
\end{bmatrix} \\
&=
\begin{bmatrix} 
a & b \\ 
0 & 0 
\end{bmatrix}
\ast
\begin{bmatrix}
x_{1,1} & x_{1,2} & \cdots & x_{1,28} \\
x_{2,1} & x_{2,2} & \cdots & x_{2,28} \\
\vdots  & \vdots  & \ddots & \vdots  \\
x_{28,1} & x_{28,2} & \cdots & x_{28,28}
\end{bmatrix} -
\begin{bmatrix} 
0 & 0 \\ 
-c & -d 
\end{bmatrix}
\ast
\begin{bmatrix}
x_{1,1} & x_{1,2} & \cdots & x_{1,28} \\
x_{2,1} & x_{2,2} & \cdots & x_{2,28} \\
\vdots  & \vdots  & \ddots & \vdots  \\
x_{28,1} & x_{28,2} & \cdots & x_{28,28}
\end{bmatrix}.
\end{align*}
In this way, the kernels are positive and could be processed by direct detection in our optical hardware. E/O modulators and waveshapers are used to encode the kernels into the optical carrier.
The positive and negative kernels could also be processed in one time round by setting the intermediate value, which is \[\frac{P_{\text{max}} + P_{\text{min}}}{2}\] as the reference level, in this way, the \( P_{\text{maxpd}}\) corresponds to floating point value 1 and \( P_{\text{minpd}}\) corresponds to floating point value -1, any values between -1 and 1 could be extracted using the methods described in ~\eqref{eq:voltage_extraction} and ~\eqref{eq:output_voltage}.

\subsection{Subsection 2: Electrical configurable E/O modulator for convolution operation }
\label{sec:s42}
\begin{figure}[H]
    \centering
    \includegraphics[width=0.6\textwidth]{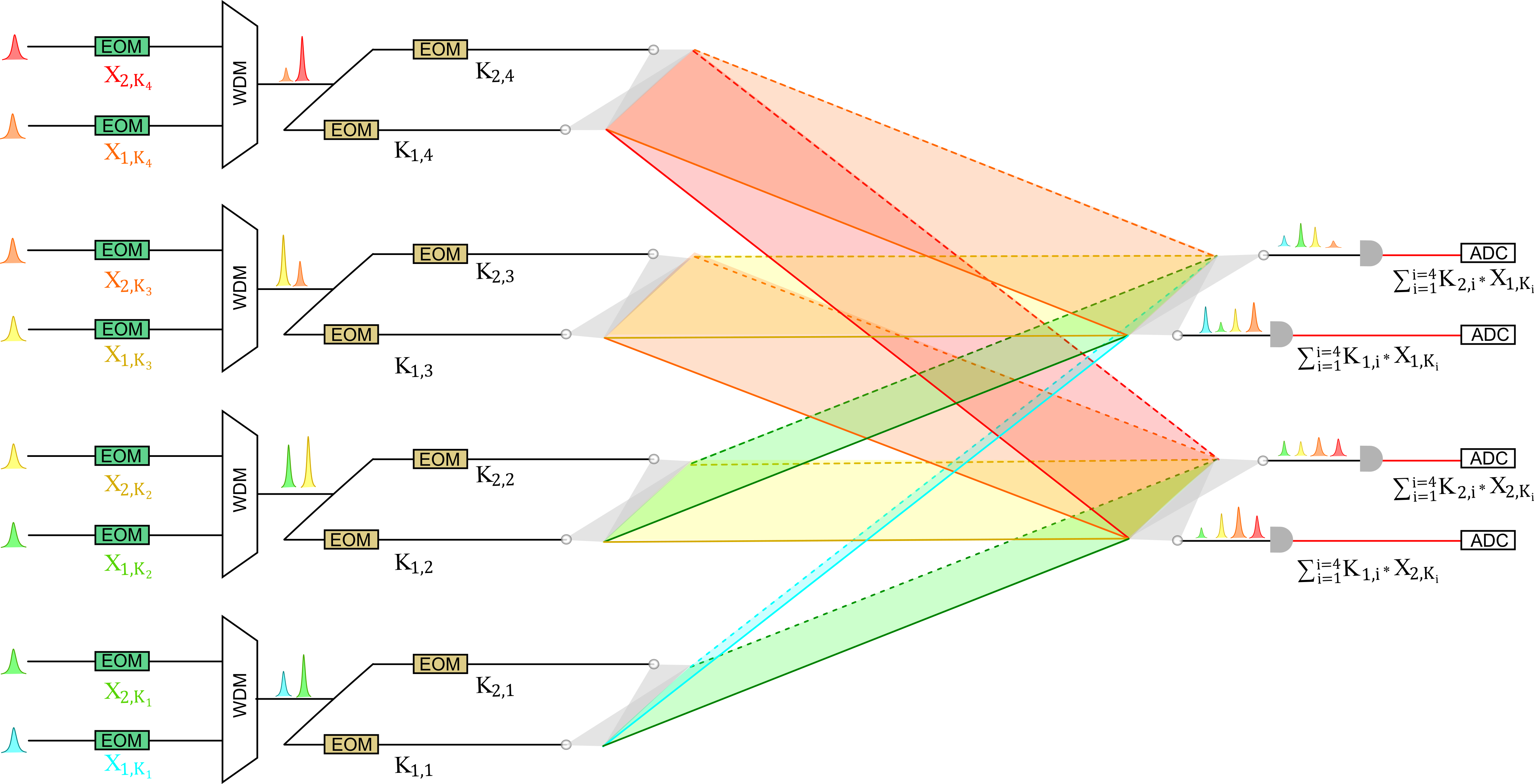}
    \caption{Convolution setup of mapping two \( 2*2\) kernels into the optical hardware, eight kernel modulators and eight data modulators are used to encode the kernel and the data.}
    \label{fig:single_step_conv_setup}
\end{figure}

\begin{figure}[H]

    \centering
    \includegraphics[width=0.6\textwidth]{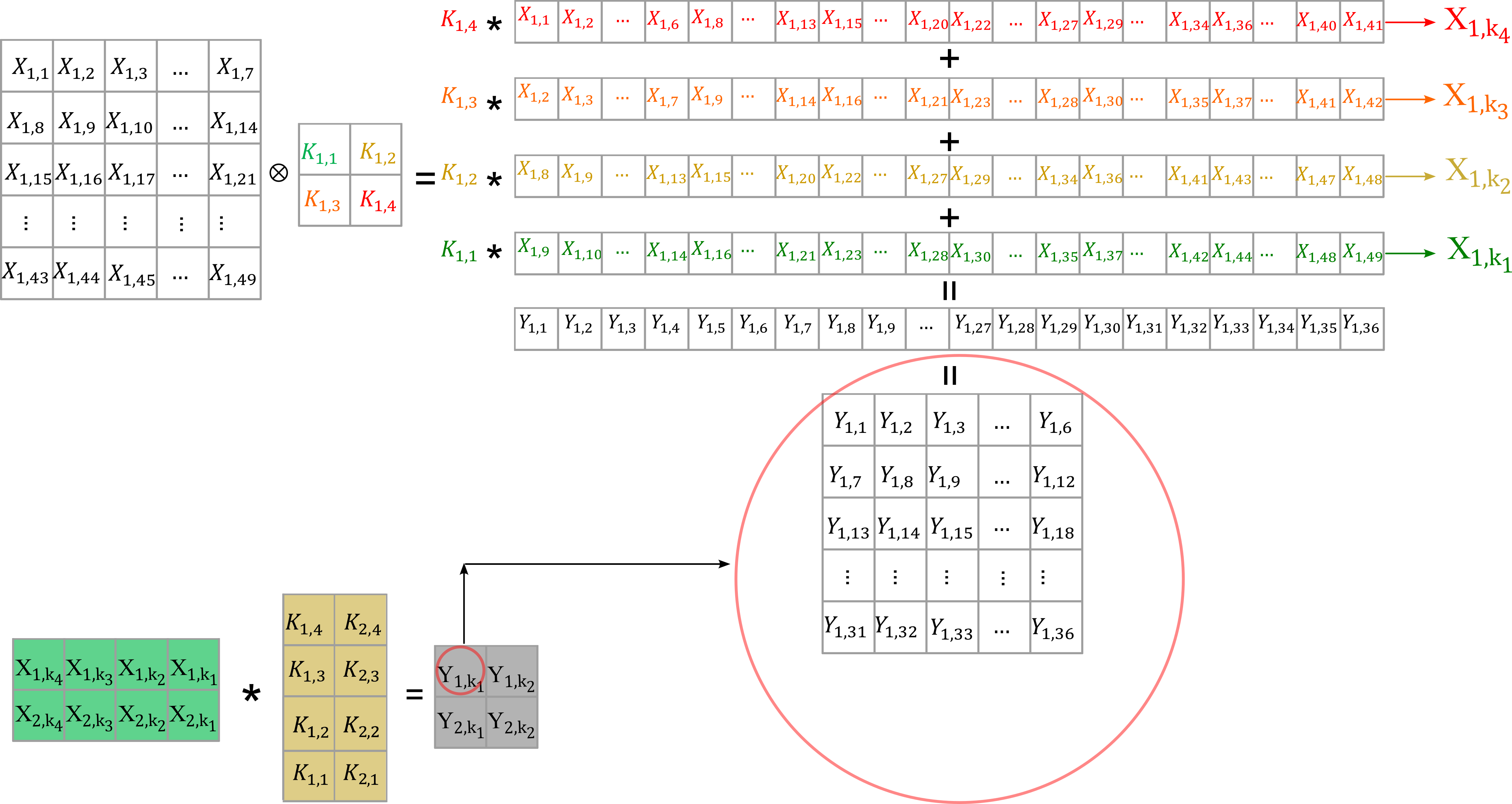}
    \caption{Processing the data matrix for \(2*2\) kernel convolution, the data matrix is processed and mapped to four data modulators. }
    \label{fig:single_step_conv_data_processing}
\end{figure}

\begin{figure}[H]
    \centering
    \includegraphics[width=0.6\textwidth]{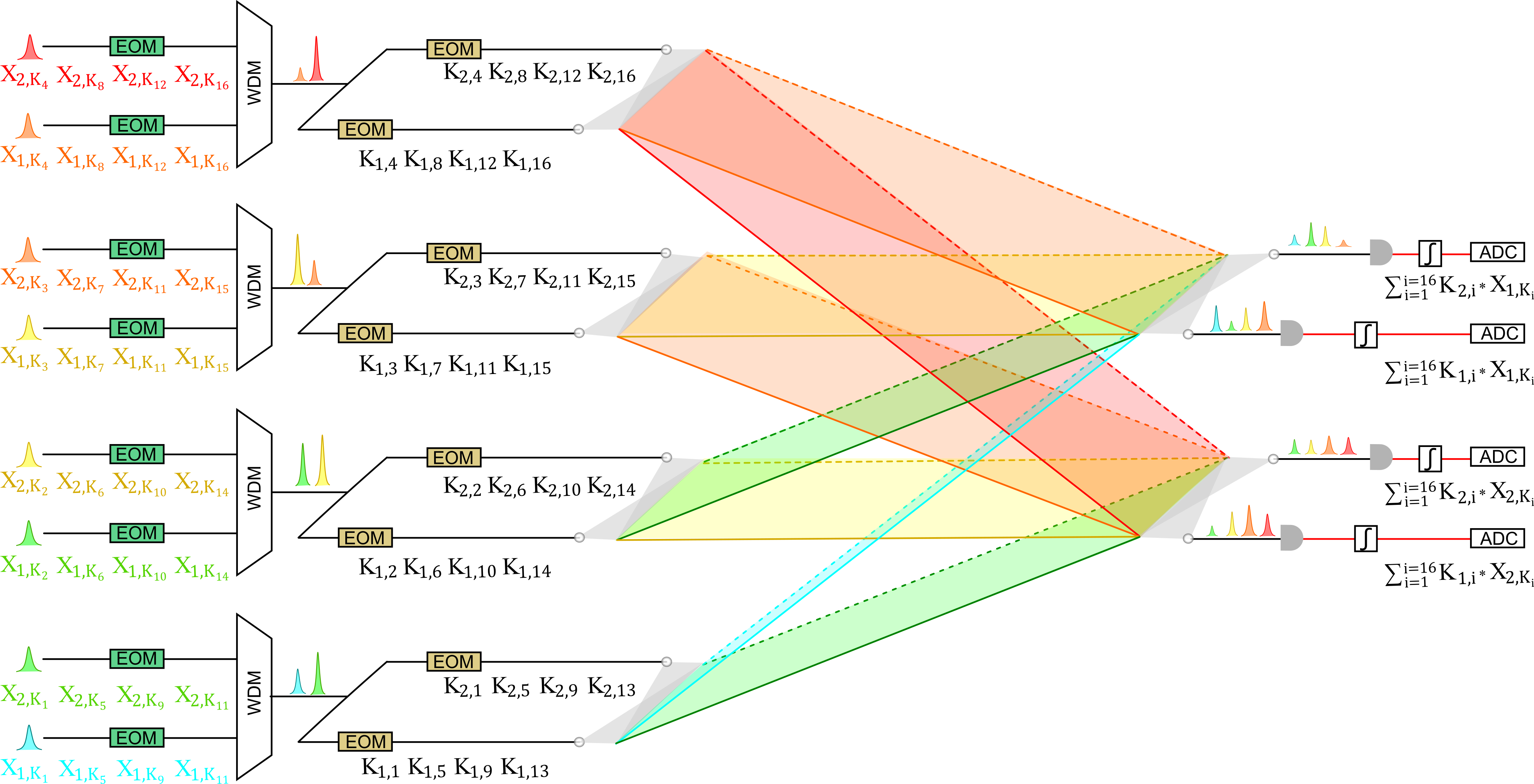}
    \caption{Convolution setup of mapping two \( 4*4\) kernels into the optical hardware, 4 modulators encodes the \( 4*4\) kernel in 4 time steps.}
    \label{fig:time_integrated_conv_setup}
\end{figure}

\begin{figure}[H]
    \centering
    \includegraphics[width=0.6\textwidth]{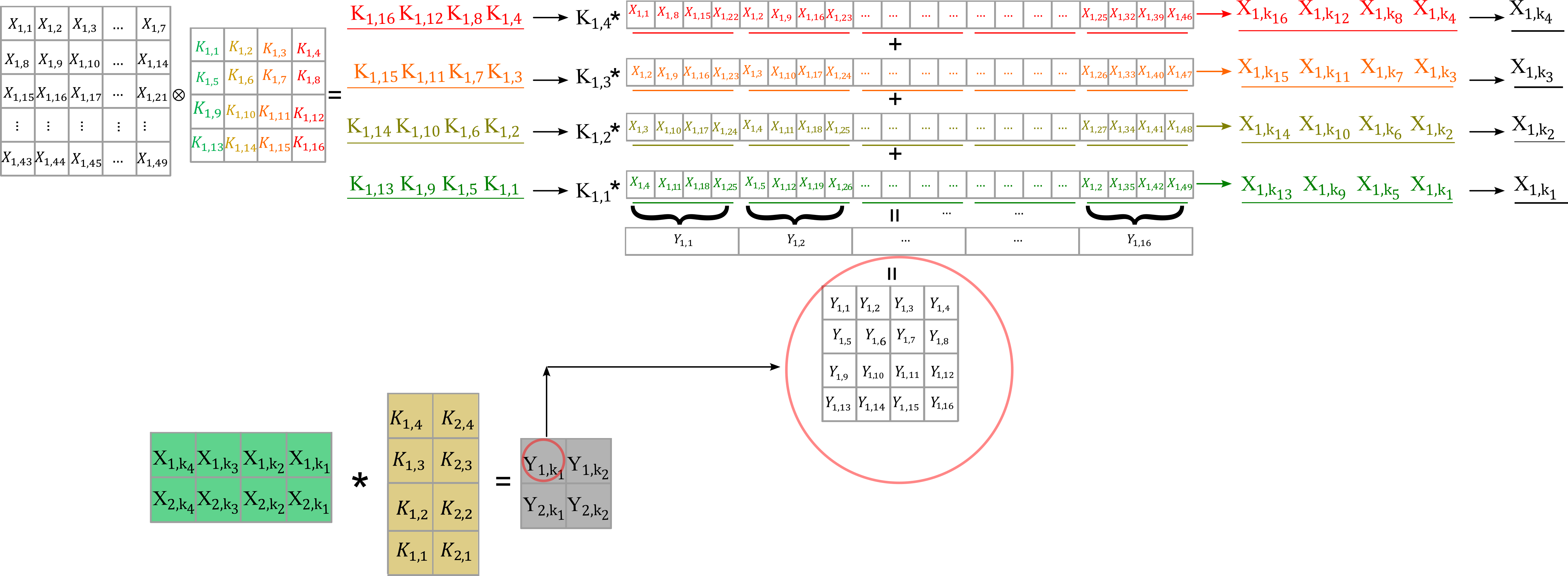}
    \caption{Processing the data matrix for \( 4*4\) kernel convolution, the data matrix is processed and mapped to four data modulators in 4-time steps. }
    \label{fig:time_integrated_conv_data_processing}
\end{figure}

\begin{figure}[H]
    \centering
    \includegraphics[width=0.8\textwidth]{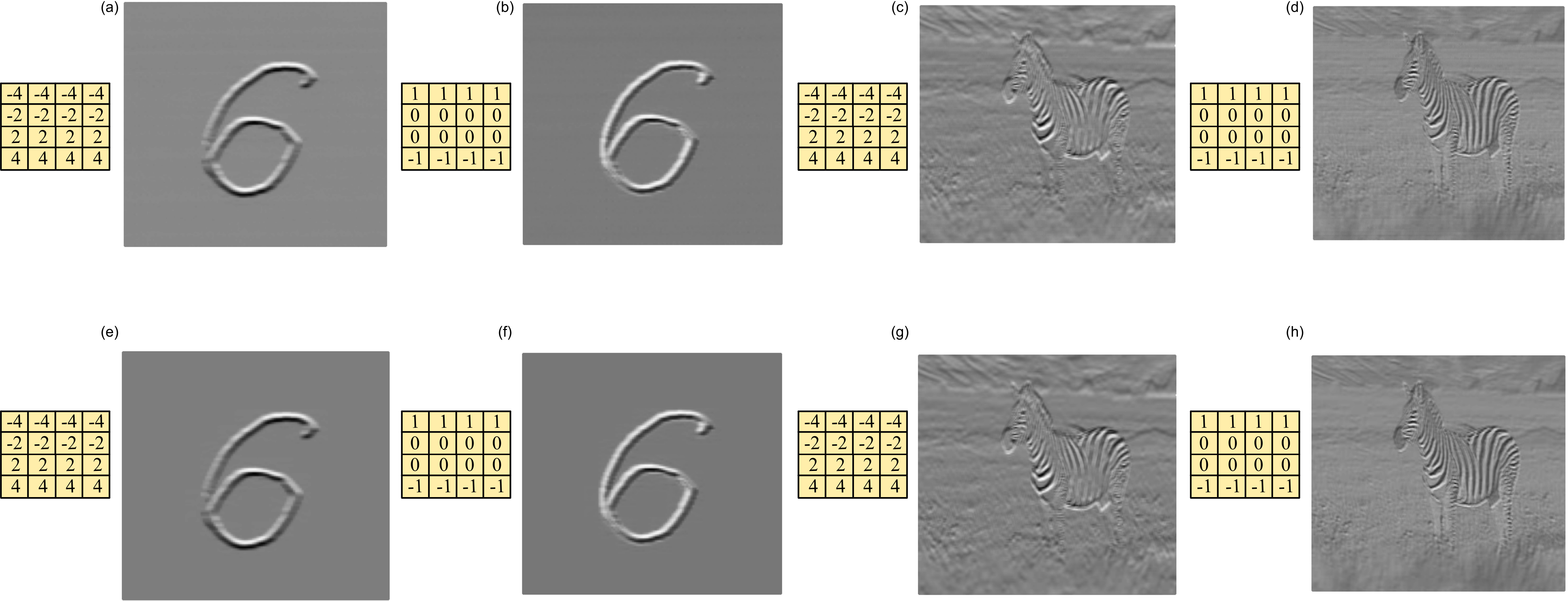}
    \caption{The convolution results for the \( 4*4\) kernel with image matrix using analog time integrator, the top 4 figures are the convolution images using the optical hardware, where the bottom figures are digital convolution results..}
    \label{fig:time_integrated_conv_result}
\end{figure}

Fig.~\ref{fig:single_step_conv_setup} shows the setup of the convolution operation. Two \( 2*2\) kernels are encoded through eight E/O intensity modulators, and each weight modulator encodes one element of these two kernels, in this way, the output value and input voltage for each modulator are constant values and the convolution operation for a single patch could be finished in a single time step.
The image data matrix needs to be processed before it could be mapped to the data modulators, Fig.~\ref{fig:single_step_conv_data_processing} shows how to process the image matrix data for convolution.
Using the analog time integrator, we can map larger kernels into the optical hardware. Fig.~\ref{fig:time_integrated_conv_setup} and Fig.~\ref{fig:time_integrated_conv_data_processing} show the setup and the data processing steps when using the optical hardware to perform a \( 4*4\) kernel convolution, analog time integrator accumulate the multiplication every four-time steps, Fig.~\ref{fig:time_integrated_conv_result} shows the convolution outputs of the image with two \( 4*4\) kernels using the analog time integrator, the optical convolutional results matches well with the digital convolution results, indicating high accuracy of the grating beam routing system.

\newpage
\subsection{Subsection 3: Programmable waveshapers for convolution operation}
\label{sec:s43}
\begin{figure}[H]
    \centering
    \includegraphics[width=0.9\textwidth]{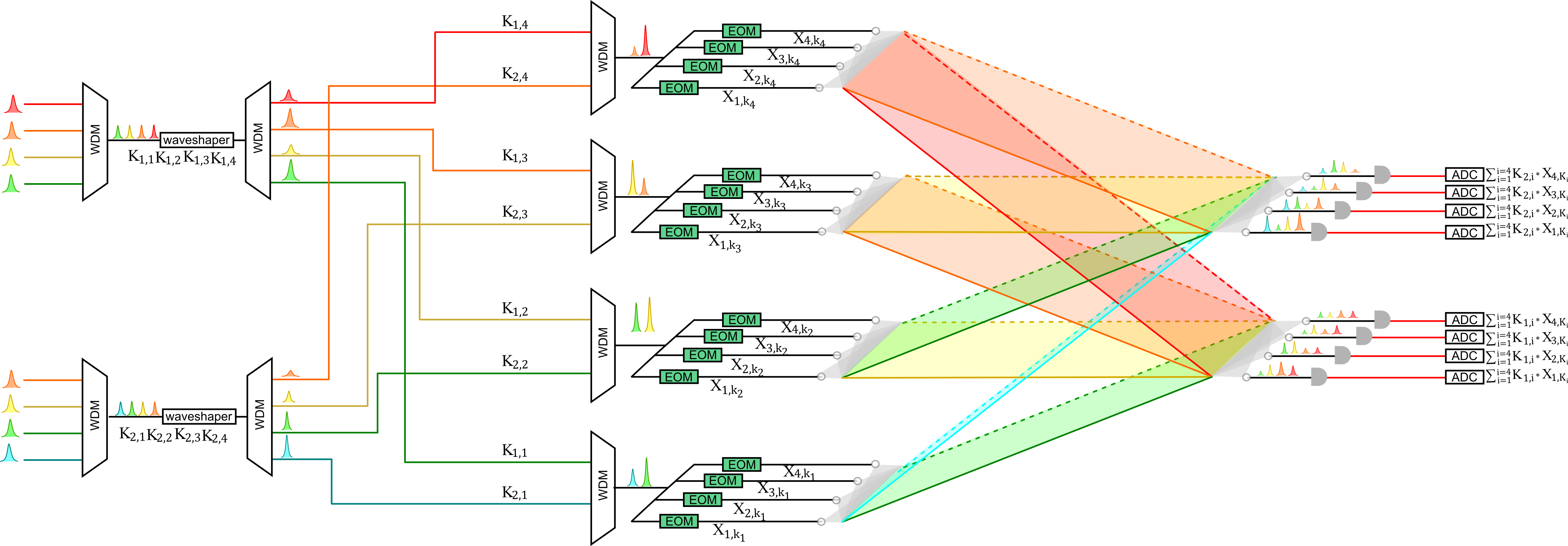}
    \caption{The convolution setup for the \( 2*2\) kernel with image matrix using waveshaper.}
    \label{fig:waveshaper_conv_setup}
\end{figure}

Programmable waveshapers are also used to perform the convolution operation, compared with the E/O modulator, waveshaper could provide a much larger on-off ratio and extinction ratio of the signal.
Fig.~\ref{fig:waveshaper_conv_setup} shows the setup and results of the convolution operation, two \( 2*2\) kernels and 4 images are processed simultaneously, the optical convolution results match well with the digital convolution results across multiple wavelengths and spatial channels, which further verifies the high parallelism, low crosstalk and high accuracy of the optical beam routing system.
The waveshaper switching speed is low, so we didn't perform the large-size kernel convolution using waveshaper.  
\begin{figure}[H]
    \centering
    \includegraphics[width=0.9\textwidth]{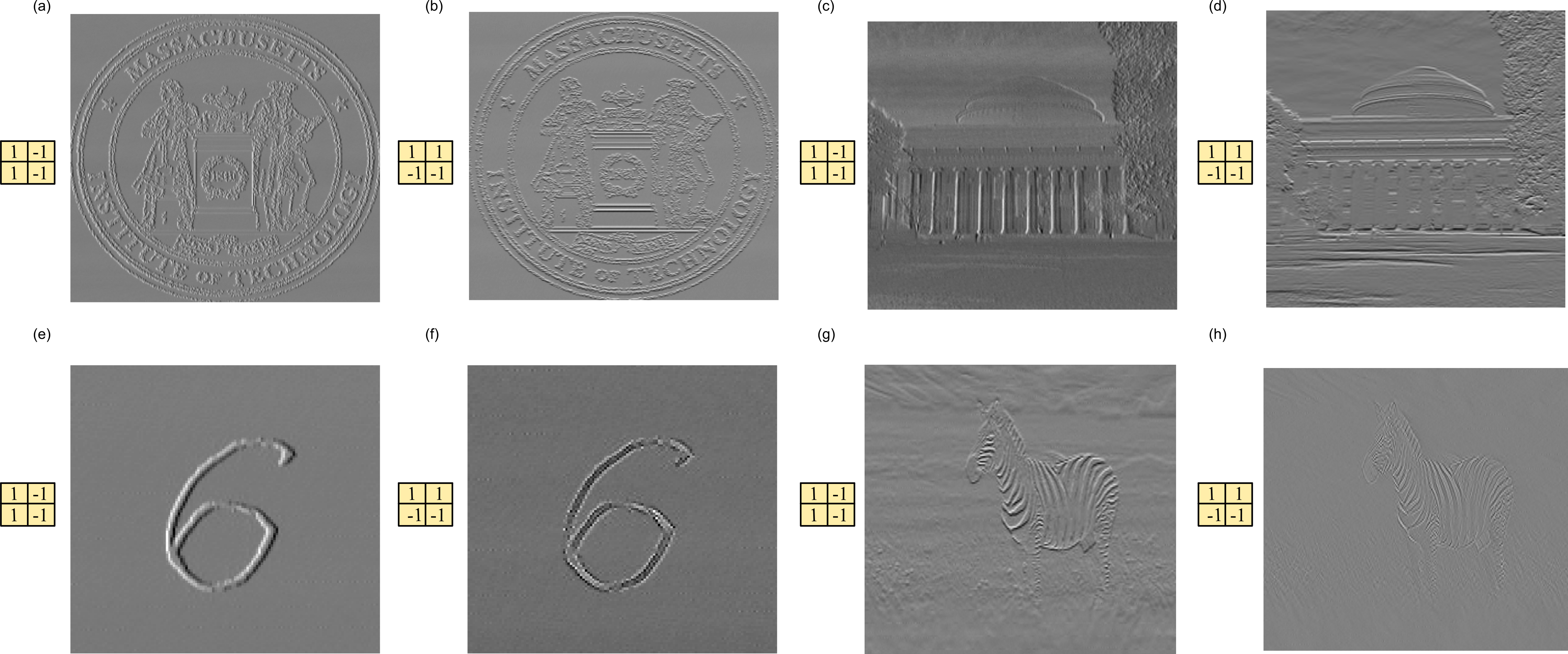}
    \caption{The convolution results for the \( 2*2\) kernel with image matrix using waveshaper.}
    \label{fig:waveshaper_conv_result}
\end{figure}

\newpage
\section{Optical parallelism and energy efficiency}
\label{sec:s51}
The grating beam routing system uses SDM, wavelength Mux, wavelength DeMux, free-space beam routing system, and analog time integrator to achieve low energy, high parallelism operation. The energy consumption of the system is shown in Table~\ref{tab:2}, where N and M are the fiber array channels, T is the analog time integrator integration period.

\begin{table}[!htbp]
    \centering
    \vspace{0.75em} 
    \textbf{Energy consumption of the grating beam routing system based optical hardware.}
    \vspace{0.5em} 
    \label{tab:2}
    \begin{tabular}{|>{\centering\arraybackslash}m{3.5cm}|>{\centering\arraybackslash}m{3cm}|>{\centering\arraybackslash}m{3cm}|>{\centering\arraybackslash}m{4cm}|>{\centering\arraybackslash}m{3cm}|}
        \hline
        \textbf{Components} & 
        \textbf{Energy per Operation} & 
        \textbf{MACs per Operation} & 
        \textbf{Energy per MAC} & 
        \textbf{Number of Components} \\ 
        \hline
       Data modulator & $\sim 1\,\text{pJ}$ & $M$ & $1/M\,\text{pJ}$ & $N^2$ \\ 
       Weight modulator & $\sim 1 \, \text{pJ}$ & $N$ & $1/N\,\text{pJ}$ & $NM$ \\ 
       DAC for data modulator & $\sim 1 \, \text{pJ}$ & $M$ & $1/M\,\text{pJ}$ & $N^2$ \\ 
       DAC for weight modulator & $\sim 1 \, \text{pJ}$ & $N$ & $1/N\,\text{pJ}$ & $NM$ \\ 
       ADC & $\sim 1 \, \text{pJ}$ & $NT$ & $1/(NT)\,\text{pJ}$ & $NM$ \\  
       Photoreceiver & $\sim 1 \, \text{fJ}$ & $N$ & $1/N\,\text{fJ}$ & $NM$ \\  
       Analog integrator & $\sim 1 \, \text{fJ}$ & $NT$ & $1/(NT)\,\text{fJ}$ & $NM$ \\ 
       Nonlinearity & $< 100 \, \text{fJ}$ & $NT$ & $100/(NT)\,\text{fJ}$ & $NM$ \\ 
       \hline
       Total & $-$ & $-$ & $(2/M+2/N+1/NT)\,\text{pJ}$ & $-$ \\ 
        \hline
    \end{tabular}
\end{table}

\begin{table}[!htbp]
    \centering
    \vspace{0.75em} 
    \textbf{Energy consumption of the grating beam routing system with optimized optical DAC.}
    \vspace{0.5em} 
    \label{tab:3}
    \begin{tabular}{|>{\centering\arraybackslash}m{3.5cm}|>{\centering\arraybackslash}m{3cm}|>{\centering\arraybackslash}m{3cm}|>{\centering\arraybackslash}m{4cm}|>{\centering\arraybackslash}m{3cm}|}
        \hline
        \textbf{Components} & 
        \textbf{Energy per Operation} & 
        \textbf{MACs per Operation} & 
        \textbf{Energy per MAC} & 
        \textbf{Number of Components} \\ 
        \hline
       ODAC for data encoding & $\sim 40 \, \text{fJ}$ & $M$ & $40/M\,\text{fJ}$ & $N^2$ \\ 
       ODAC for weight encoding & $\sim 40 \, \text{fJ}$ & $N$ & $40/N\,\text{fJ}$ & $NM$ \\ 
       ADC & $\sim 1 \, \text{pJ}$ & $NT$ & $1/(NT)\,\text{pJ}$ & $NM$ \\  
       Photoreceiver & $\sim 1 \, \text{fJ}$ & $N$ & $1/N\,\text{fJ}$ & $NM$ \\  
       Analog integrator & $\sim 1 \, \text{fJ}$ & $NT$ & $1/(NT)\,\text{fJ}$ & $NM$ \\ 
       Nonlinearity & $< 100 \, \text{fJ}$ & $NT$ & $100/(NT)\,\text{fJ}$ & $NM$ \\ 
       \hline
       Total & $-$ & $-$ & $(40/M+40/N+1101/NT+1/N)\,\text{fJ}$ & $-$ \\ 
        \hline
    \end{tabular}
\end{table}

For an optical neural network architecture without using the above-mentioned technique, the system performance is shown in Table~\ref{tab:4}. In this case, the energy consumption is $5\,\text{pJ}$ per MAC. 
\begin{table}[!htbp]
    \centering
    \vspace{0.75em} 
    \textbf{Energy consumption of optical hardware without using grating beam routing, optical fanout, and time integration.}
    \vspace{0.5em} 
    \label{tab:4}
    \begin{tabular}{|>{\centering\arraybackslash}m{3.5cm}|>{\centering\arraybackslash}m{3cm}|>{\centering\arraybackslash}m{3cm}|>{\centering\arraybackslash}m{3cm}|>{\centering\arraybackslash}m{3cm}|}
        \hline
        \textbf{Components} & 
        \textbf{Energy per Operation} & 
        \textbf{MACs per Operation} & 
        \textbf{Energy per MAC} & 
        \textbf{Number of Components} \\ 
        \hline
       Data modulator & $\sim 1\,\text{pJ}$ & $1$ & $1\,\text{pJ}$ & $N^2M$ \\  
       Weight modulator & $\sim 1 \, \text{pJ}$ & $1$ & $1\,\text{pJ}$ & $N^2*M$ \\ 
       ADC for data modulator & $\sim 1 \, \text{pJ}$ & $1$ & $1\,\text{pJ}$ & $N^2*M$ \\ 
       ADC for weight modulator & $\sim 1 \, \text{pJ}$ & $1$ & $1\,\text{pJ}$ & $N^2*M$ \\ 
       DAC & $\sim 1 \, \text{pJ}$ & $1$ & $1\,\text{pJ}$ & $N^2*M$ \\  
       Photoreceiver & $\sim 1 \, \text{fJ}$ & $1$ & $1\,\text{fJ}$ & $N^2*M$ \\  
       Analog integrator & $\sim 1 \, \text{fJ}$ & $1$ & $1\,\text{fJ}$ & $N^2*M$ \\ 
       Nonlinearity & $< 100 \, \text{fJ}$ & $1$ & $100\,\text{fJ}$ & $N^2*M$ \\ 
       \hline
       Total & $-$ & $-$ & $5\,\text{pJ}$ & $-$ \\ 
        \hline
    \end{tabular}
\end{table}

\section{CNNs and DNNs for image classification}

\subsection{Subsection 1: Digital training of the network}
\label{sec:s61}
\begin{figure}[t!]
    \centering
    \includegraphics[width=0.4\textwidth]{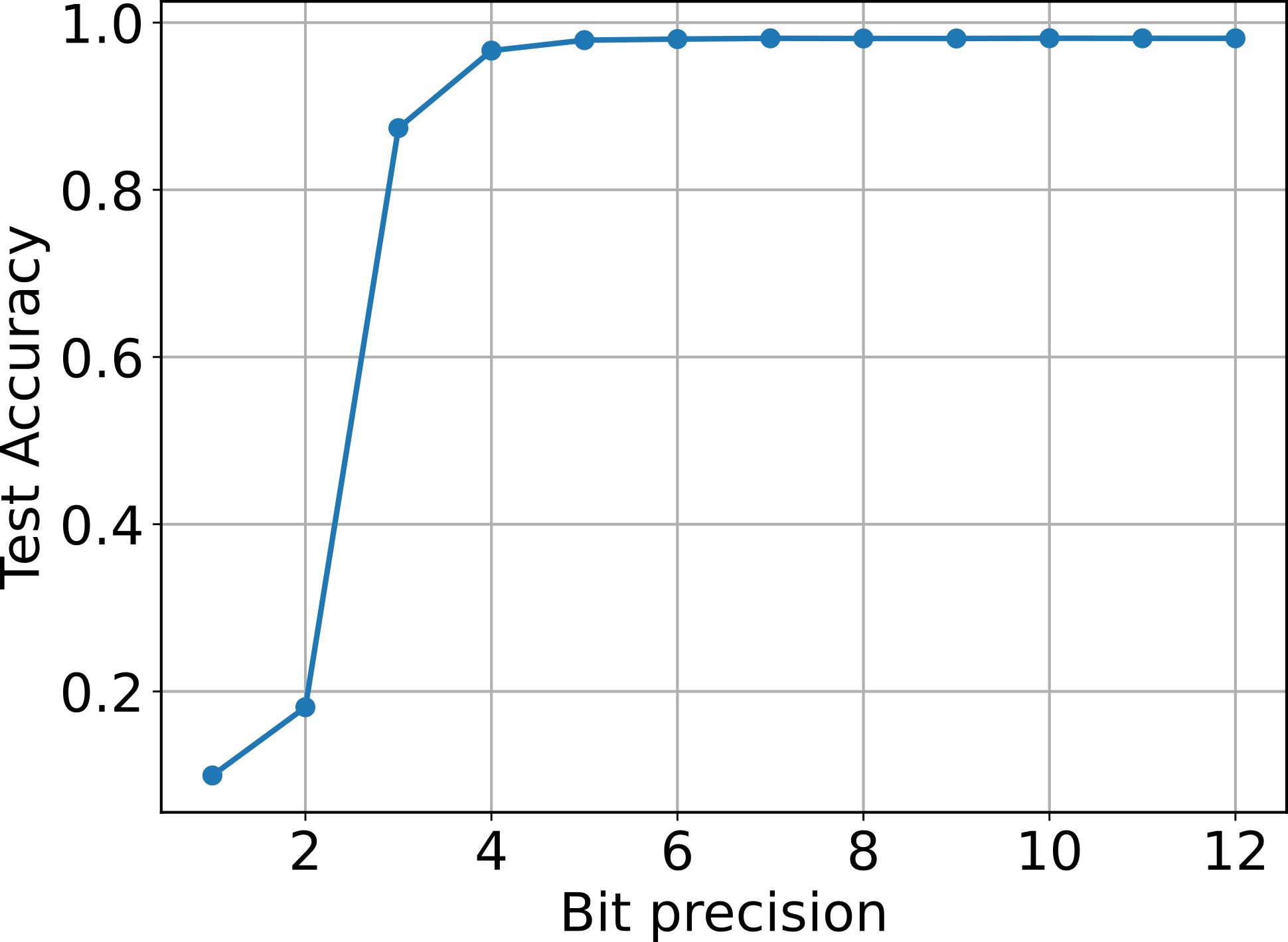}
    \caption{The convolution results for the \( 2*2\) kernel with image matrix using waveshaper.}
    \label{fig:network accuracy versus bit precision}
\end{figure}

The CNN and FC NN were trained on a standard digital computer in Python with the PyTorch library on 50,000 training images for the MNIST and Fashion-MNIST datasets. 10,000 images were reserved for validation sets to fine-tune the network hyperparameters and optical setup. Each dataset was normalized by its standerd deviation, and L2 regularization with $L_2 = 0.0001$ and $10\%$ dropout were applied to each layer. Gaussian noise was also added to each activation value at every layer, with a standard deviation of $0.25 * \delta$, where $\delta$ is the standard deviation of the activation value across a batch. The nonlinear activation function on the final layer was SoftMax. The Adam optimizer minimized the categorical cross-entropy loss function with a batch size of 100 and learning rate of 0.001.
Since we use the noise aware training and the quantization aware training, the actual network is very robust to the noise and suitable for low bit precision tasks. Fig.~\ref{fig:network accuracy versus bit precision} shows the network testing accuracy for data with different bit precisions.

\subsection{Subsection 2: Implementation of the optical CNNs and DNNs}
\label{sec:s71}
 We perform a benchmark one-layer CNN and a following fully connected DNN inferences using the optical processor. We apply four \(2\times 2\) kernels to extract the features of these images, each input image, with \(28\times 28\) pixels, is encoded in 729 time-steps to four data modulators .
The grating beam routing optical processor performs parallel batch operations, two \(2\times 2\) kernels, and two MNIST images are convolved by the grating beam routing processor simultaneously. The outputs from the convolutional layer, which contain 2,916 pixels of 1,000 images, are detected by amplified photodetectors. Series of operations like decoding the detector data from voltages to analog values, ReLU nonlinear activations, and data normalizations are performed digitally. The CNN outputs are fed into a fully connected DNNs, implemented by the same optical system.

The DNNs have one input layer (2,916 neurons), one fully connected hidden layer (100 neurons), and one output layer (10 neurons). These 2,916-pixel values are mapped to four data modulators in 729 time-steps, the first two columns of the weight matrix are mapped to the eight weight modulators, with each weight column encoded by four weight modulators in 729 time-steps, and this process is repeated for all the 100 weight matrix columns. The fully connected hidden layer output values are recorded by analog time integrators. This is a batch process, and the same operations are performed for two input images simultaneously. Decoding, nonlinear ReLU activation, and normalization are performed digitally and the output results are sent to the system to perform the \(100\times 10\) matrix multiplication. Similar batch operations are performed by encoding the weight matrix and data matrix to the modulators in 25 time-steps. The outputs are converted into probabilities by a digital SoftMax operation.

\subsection{Subsection 3: Latency of the image classification Model}
\label{sec:s81}
We take MNIST images (each with \(28\times 28\) pixels) for consideration. For a \(N\times M\) channel fiber array, \(N\) images with \(M\) kernels, each kernel with the dimension of \(\sqrt{N}\times \sqrt{N}\) will be processed simultaneously, the accumulation of kernel elements and data pixels multiplication is achieved through WDM. The \(N\) input images are mapped to \(N^2\) data modulators, with each image being flattened to \(N(28-\sqrt{N}+1)^2\) pixels and encoded by \(N\) data modulators in \((28-\sqrt{N}+1)^2\) time steps. The \(M\) kernels (each with the dimension of \(\sqrt{N}\times \sqrt{N}\)) are encoded to \(N\times M\) weight modulators, with each modulator has constant input RF voltage and output intensity. In a single clock rate, \(N\times\ M\) pixels of \(N\times\ M\) new images are generated, \((28-\sqrt{N}+1)^2\) time steps are needed to finish the \((\sqrt{N}\times\sqrt{N})\) convolutional operation of \(M\) kernels among  \(N\) images. For a total MNIST image numbers of \(P\), the time consumption for the convolutional operation is \(\frac{P}{N}\left(28-\sqrt{N}+1\right)^2\frac{1}{C}\). 

For the subsequent DNN operation, the analog time integration is employed. Suppose the fully connected hidden layer and output layer neurons are \(D_{1}, D_{2}\). The weight matrix size is \(M(28-\sqrt{N}+1)^2\times D_{1}\) and \(D_{1}\times D_{2}\). For the hidden layer, each analog time integrator needs to accumulate \(\frac{M\left(28-\sqrt{N}+1\right)^2}{N}\) time steps over \(N\) different wavelengths before the readout, and this process needs to perform \(\frac{D_1}{M}\) times across \(M\) different spatial channels for all the \(D_{1}\) weight matrix columns. The system supports batch operation, in each time step, \(N\) images are processed by the system simultaneously. Based on this calculation, in order to process these \(P\) images, the needed time is \(\frac{D_1P\left(28-\sqrt{N}+1 \right)^2}{CN^2}\). Similarly, the time to process output layer is \(\frac{1}{C}\frac{D_2}{M}\frac{P}{N}\frac{D_1}{N}\)N. Other operations, including the ReLU nonlinear activation, SoftMax operation, are processed digitally. So the total time needed to perform the image classification is \(\frac{D_1P\left( 28 - \sqrt{N} + 1 \right)^2}{CN^2} + \frac{D_1PD_2}{CMN^2} + \frac{P\left( 28-\sqrt{N}+1\right)^2}{CN}\), while for the system with $O{(N)}$ throughput, the processing time is \( N^2M\) longer. For the current \(4\times 4\) fiber array, the time to process 1,000 MNIST images is \SI{96}{\milli\second}. The latency could be reduced by multiplexing more wavelength and spatial channels through a large channel number fiber array, a \(30\times 30\) channel fiber array with \(10\,\mathrm{GSas^{-1}}\) would reduce the processing time of 1,000 images to \SI{1140}{\nano\second}.

\newpage
\section{System picture}
\label{sec:s71}
\begin{figure}[H]
    \centering
    \includegraphics[width=0.3\textwidth]{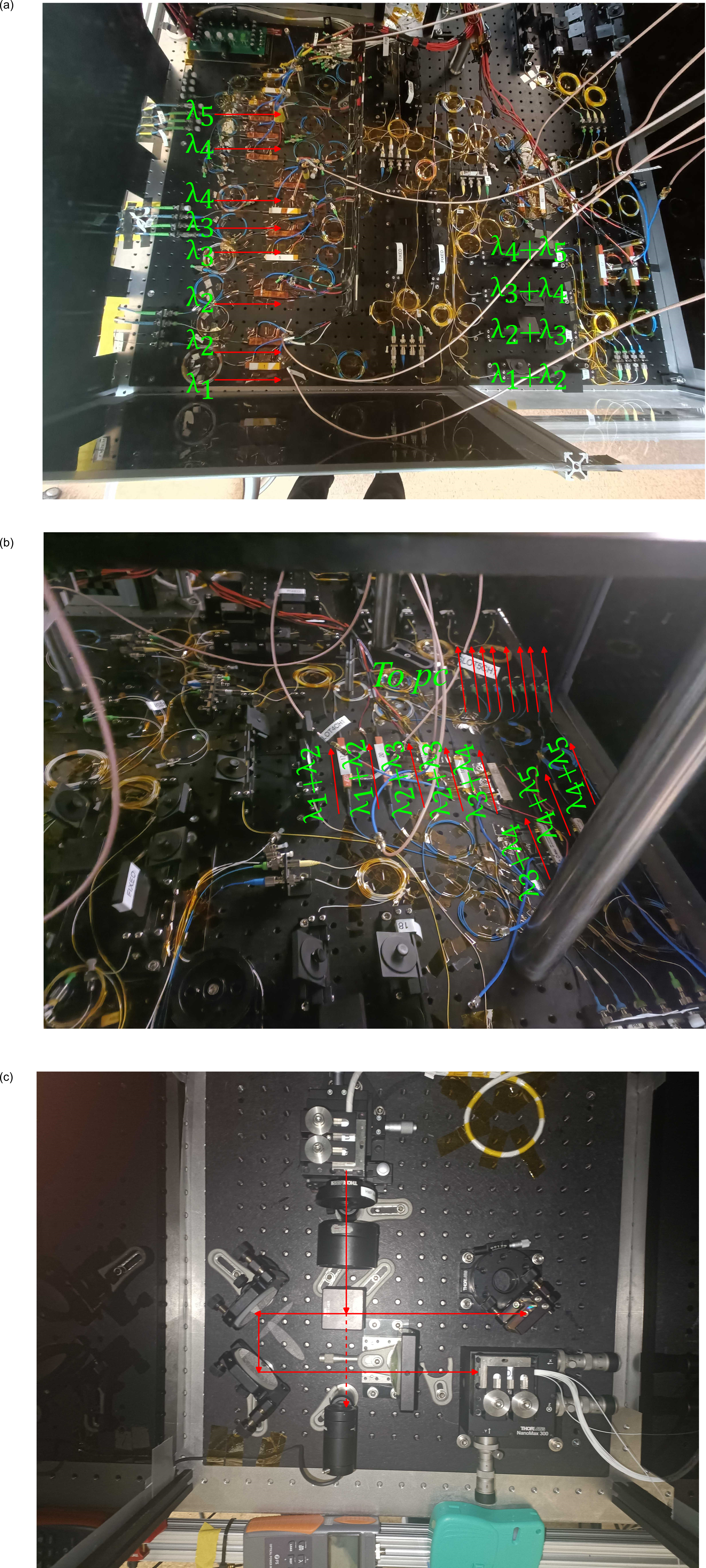}
    \caption{Figure of the builded grating optical neural network setup.}
    \label{fig:setup figure}
\end{figure}

Fig.~\ref{fig:setup figure} shows the setup figure, where from top to bottom, the data modulators, the weight modulators, and the grating beam routing system are shown.

\bibliography{main}

\end{document}